\documentclass[journal]{IEEEtran}

\usepackage{amsmath,amsfonts}
\usepackage{array}
\usepackage[caption=false,font=normalsize,labelfont=sf,textfont=sf]{subfig}
\usepackage{textcomp}
\usepackage{stfloats}
\usepackage{url}
\usepackage{verbatim}
\usepackage{graphicx}
\def\BibTeX{{\rm B\kern-.05em{\sc i\kern-.025em b}\kern-.08em
    T\kern-.1667em\lower.7ex\hbox{E}\kern-.125emX}}
\usepackage{balance}

\usepackage{cite}
\usepackage{xcolor}

\usepackage{hyperref}
\hypersetup{
    colorlinks=true,
    linkcolor=blue,
    citecolor=blue,
    filecolor=black,      
    urlcolor=black,
    }

\usepackage{algorithm}
\usepackage{algorithmic}

\usepackage{makecell}

\usepackage{multirow}

\usepackage{mathtools}
\DeclarePairedDelimiter\ceil{\lceil}{\rceil}

\usepackage{nomencl}
\usepackage{ifthen}
\renewcommand{\nomgroup}[1]{%
\ifthenelse{\equal{#1}{I}}{\item[\textbf{Index and Sets}]}{%
\ifthenelse{\equal{#1}{A}}{\item[\textbf{Abbreviations}]}{%
\ifthenelse{\equal{#1}{P}}{\item[\textbf{Parameters}]}{%
\ifthenelse{\equal{#1}{V}}{\item[\textbf{Variables}]}{}}}}
}

\begin{document}



    \title{A Stochastic Incentive-based Demand Response Program for Virtual Power Plant with Solar, Battery, Electric Vehicles, and Controllable Loads}

    \author{Pratik Harsh\IEEEauthorrefmark{1}, Hongjian Sun\IEEEauthorrefmark{2}, Debapriya Das\IEEEauthorrefmark{3}, Goyal Awagan\IEEEauthorrefmark{4}, and Jing Jiang\IEEEauthorrefmark{5}
    \thanks{\IEEEauthorrefmark{1}Corresponding Author.}
    \thanks{P. Harsh and H. Sun are with the Department of Engineering, Durham University, Durham, United Kingdom (e-mail: \IEEEauthorrefmark{1}pratik.harsh@durham.ac.uk; \IEEEauthorrefmark{2}hongjian.sun@durham.ac.uk).}
    \thanks{D. Das is with the Department of Electrical Engineering, Indian Institute of Technology, Kharagpur, West Bengal 721302, India (e-mail: \IEEEauthorrefmark{3}ddas@ee.iitkgp.ac.in).}
    \thanks{G. Awagan and J. Jiang are with the Department of Mathematics, Physics \& Electrical Engineering, Northumbria University, Newcastle, United Kingdom (e-mail: \IEEEauthorrefmark{4}goyal.awagan@northumbria.ac.uk; \IEEEauthorrefmark{5}jing.jiang@northumbria.ac.uk).}}

    \maketitle

    \begin{abstract}
        The growing integration of distributed energy resources (DERs) into the power grid necessitates an effective coordination strategy to maximize their benefits. Acting as an aggregator of DERs, a virtual power plant (VPP) facilitates this coordination, thereby amplifying their impact on the transmission level of the power grid. Further, a demand response program enhances the scheduling approach by managing the energy demands in parallel with the uncertain energy outputs of the DERs. This work presents a stochastic incentive-based demand response model for the scheduling operation of VPP comprising solar-powered generating stations, battery swapping stations, electric vehicle charging stations, and consumers with controllable loads. The work also proposes a priority mechanism to consider the individual preferences of electric vehicle users and consumers with controllable loads. The scheduling approach for the VPP is framed as a multi-objective optimization problem, normalized using the utopia-tracking method. Subsequently, the normalized optimization problem is transformed into a stochastic formulation to address uncertainties in energy demand from charging stations and controllable loads. The proposed VPP scheduling approach is addressed on a 33-node distribution system simulated using MATLAB software, which is further validated using a real-time digital simulator.
    \end{abstract}

    
    
    \makenomenclature
    
    \nomenclature[I]{$\mathcal{S}^T,t$}{Set and index denoting operating intervals.}
    \nomenclature[I]{$\mathcal{S}^N,n$}{Set and index denoting nodes of distribution network at which solar station, charging station (CS), and battery swapping station (BSS) are integrated.}
    \nomenclature[I]{$\mathcal{S}_V^n,v$}{Set and index denoting electric vehicles (EVs) at CS of $n^{th}$ node.}
    \nomenclature[I]{$\mathcal{S}_B^n,b$}{Set and index denoting batteries at BSS of $n^{th}$ node.}
    \nomenclature[I]{$\mathcal{S}^L$}{Set of nodes integrated with controllable loads.}
    \nomenclature[I]{$\mathcal{S}^U,u$}{Set and index denoting uncertain input variables.}
    
    \nomenclature[P]{$P_{s0}$}{Nominal power rating of a solar module (in kWp).}
    \nomenclature[P]{$V_{MPP},I_{MPP}$}{Solar module voltage (in V) and current (in A) at its maximum power point.}
    \nomenclature[P]{$V_{OC},I_{SC}$}{Open circuit voltage (in V) and short circuit current (in A) of a solar module.}
    \nomenclature[P]{$k_{v},k_{i}$}{Voltage (in \%V/$^0$C) and current (in \%A/$^0$C) temperature coefficients.}
    \nomenclature[P]{$T_{o/a}$}{Nominal operating/ambient temperature (in $^0$C).}
    \nomenclature[P]{$s^t$}{Solar irradiance at the $t^{th}$ operating interval (in kW/m$^2$).}
    \nomenclature[P]{$\eta_{G2V/V2G}$}{EV charger efficiency during grid-to-vehicle (G2V)/vehicle-to-grid (V2G) mode.}
    \nomenclature[P]{$z_{v/b}^{n,t}$}{Binary indicator for $v^{th}$ EV or $b^{th}$ battery at $t^{th}$ operating interval connected to CS/BSS installed at $n^{th}$ node (+1, if EV/battery is in G2V/G2B mode; 0, if EV/battery is in V2G/B2G mode).}
    \nomenclature[P]{$t_{start/stop}^{n}$}{Start/stop time of load curtailment specified by consumer connected to $n^{th}$ node.}
    \nomenclature[P]{$E_{v/b}^{n}$}{Battery energy capacity of $v^{th}$ EV or $b^{th}$ battery connected to CS/BSS installed at $n^{th}$ node (in kWh).}
    \nomenclature[P]{$\overline{SoC}/\underline{SoC}$}{Maximum/minimum state of charge (SoC) level of batteries with EVs and BSS.}
    \nomenclature[P]{$t_{dep,v}^n$}{Departure time of $v^{th}$ EV connected to CS installed at $n^{th}$ node.}
    \nomenclature[P]{$\overline{P_{v/b}^n}$}{Rated power of EV/battery at CS/BSS of $n^{th}$ node (in kW).}
    \nomenclature[P]{$a_d,b_d$}{Coefficients of battery degradation.}
    \nomenclature[P]{$c_b$}{Battery investment cost (in \$/kWh).}
    \nomenclature[P]{$\eta_{G2B/B2G}$}{Battery charger efficiency during grid-to-battery (G2B)/battery-to-grid (B2G) mode.}
    \nomenclature[P]{$t_{arr,b}^n$}{Arrival time of EV registered for swapping $b^{th}$ battery connected to BSS installed at $n^{th}$ node.}
    \nomenclature[P]{$P_0^{n,t}$}{Nominal power demand of consumer connected to $n^{th}$ node at $t^{th}$ operating interval (in kW).}
    \nomenclature[P]{$z^{n,t}$}{Binary indicator to represent energy consumption from $n^{th}$ node at $t^{th}$ operating interval (+1, if energy is consumed; 0, if energy is delivered).}
    \nomenclature[P]{$c_{p/o}$}{Annualized installation/maintenance cost coefficients of solar generating station (in \$/kWp-year).}
    \nomenclature[P]{$r$}{Interest rate on investments for solar station.}
    \nomenclature[P]{$L^t$}{Lifetime of solar generating station (in years).}
    \nomenclature[P]{$w_i$}{Weighting factor accounting for the weightage of the individual objective function.}
    \nomenclature[P]{$DB$}{Daily budget of distribution system (DS) operator (in $\$$).}
    \nomenclature[P]{$m$}{Number of uncertain input variables.}
    \nomenclature[P]{$\mu_u,\sigma_u$}{Mean and standard deviation of the $u^{th}$ uncertain input variable.}
    \nomenclature[P]{$\alpha^t$}{Market electricity price rate for power purchased from the utility grid (in \$/kWh).}
    \nomenclature[P]{$T$}{Total number of operating intervals.}
    \nomenclature[P]{$\lambda_{1/2}$}{Limiting factors for load curtailments and solar generations, respectively.}
    
    \nomenclature[V]{$P_{s1}^{n}$}{Solar station capacity installed at node $n$ (in kWp).}
    \nomenclature[V]{$P_{v/b}^{n,t}$}{Power demand of $v^{th}$ EV or $b^{th}$ battery at $t^{th}$ interval connected to CS/BSS installed at $n^{th}$ node (in kW).}
    \nomenclature[V]{$\gamma^{t}$}{Incentivised electricity price rate (in \$/kWh).}
    \nomenclature[V]{$x^{n,t}$}{Load curtailment of controllable loads connected to $n^{th}$ node at $t^{th}$ operating interval (in kW).}
    \nomenclature[V]{$P_{grid}^t$}{Power delivery by the utility grid at $t^{th}$ operating interval (in kW).}
    \nomenclature[V]{$P_{inj}^{n,t}$}{Power injection by the solar-powered battery-integration CS integrated to $n^{th}$ node at $t^{th}$ operating interval (in kW).}

    \printnomenclature

    \section{Introduction}
    \label{S:Introduction}
        
        \IEEEPARstart{T}{he} increasing concerns regarding climate change are driving the global trend towards decarbonization. However, despite a significant decline in global carbon emissions during the COVID-19 pandemic, emissions from energy sources promptly surged to their highest recorded levels by the end of 2021 \textbf{\cite{2022:WEC}}. The primary barrier challenging the modern power grid is shifting towards emission-less generation while ensuring reliable access to affordable energy. To enhance cleaner production and optimize energy consumption, modern power grids have been integrated with distributed energy resources (DERs) \textbf{\cite{2016:NERC}} and demand-side control and communications \textbf{\cite{2010:Han}}. This integration empowers passive consumers to transition into proactive participants, referred to as prosumers \textbf{\cite{2018:Zafar}}, who actively oversee energy generation, storage, and consumption. However, individual prosumers exert minimal influence at the transmission level, and the intricate processing and communication infrastructure required for their market participation entails transaction costs that surpass potential gains. Additionally, fluctuations in market prices pose risks that are challenging to mitigate locally.

        Effectively managing and coordinating DERs can yield various advantages for the power grid, including diminished network losses, alleviated transmission line congestion, peak demand reduction, heightened flexibility, and improved system resilience \textbf{\cite{2020:Fernando, 2018:Helena, 2016:Christakou}}. The concept of a virtual power plant (VPP), an aggregator of DERs including renewable sources, energy storage units, EVs, and controllable loads, has emerged to better coordinate the DERs enhancing their controllability, visibility, and influence in the transmission network \textbf{\cite{2024:Xie}}. Using a blend of hardware and software, VPPs not only facilitate access to a previously untapped utility-scale behind-the-meter energy supply but also effectively manage geographically dispersed and diverse DERs to create unified demand-responsive assets \textbf{\cite{2022:Shah}}.

        Various approaches have been suggested for coordinating DERs in VPPs, broadly categorized as direct and indirect approaches \textbf{\cite{2018:Thomas}}. The direct approach allows the VPP operator to schedule DERs based on their operational requirements and users' priorities, thus ensuring predictable DER capacity and response and facilitating services requiring faster control, such as frequency regulation \textbf{\cite{2023:Yi,2022:Huishi,2020:Zhong}}. However, the processing and communication infrastructure needed for direct control might not be feasible for large VPPs. Moreover, the security and privacy issues associated with the direct methods raise significant concerns \textbf{\cite{2021:Sampath}}. In an indirect approach, the individual prosumers have the freedom to schedule their flexible loads based on their priorities and the incentive signals sent by the VPP operator \textbf{\cite{2023:Deng,2022:Wang,2012:Heussen}}. The indirect method can be employed by utilizing one-way signals, thereby diminishing communication needs and addressing privacy considerations. However, uniform pricing across multiple prosumers may lead to them all adjusting their energy consumption to coincide with low-price periods, potentially exacerbating demand peaks \textbf{\cite{2013:Dimitrios}}. Additionally, employing indirect price-based coordination strategies could amplify demand fluctuations and compromise system stability \textbf{\cite{2012:Mardavij}}. Indirect coordination strategies prove advantageous when the expenses associated with communication infrastructure outweigh the benefits of a direct approach or when prosumers are reluctant to provide direct access to the VPP operator \textbf{\cite{2016:Kostas}}.

        Furthermore, researchers have taken pioneering steps to integrate EV charge scheduling into VPP operations, aiming to enhance the support for sustainable developments. The widespread adoption of EVs presents new hurdles concerning grid integration and power system management. Uncoordinated EV charging patterns may cause significant spikes in electricity usage, potentially straining the power infrastructure, escalating costs, and compromising reliability \textbf{\cite{2018:Matteo}}. Coordinated charging strategies have surfaced as efficient methods to mitigate the adverse effects of EVs on the distribution system (DS) \textbf{\cite{2023:Harsh}}. Previous research in \textbf{\cite{2021:Saleh,2020:Yang,2020:Sheidaei}} has significantly advanced the integration of EVs into VPP operations as charging loads of DS. However, in practical scenarios, CSs serve as primary EV charging infrastructure providers \textbf{\cite{2021:Niti}}, which help to address the challenges associated with integrating EVs as an individual unit in the DS. CSs possess the capacity to manage a larger quantity of charging units compared to individual household chargers, making them potential hubs for EV energy scheduling. Consequently, initial efforts have been undertaken to integrate CSs into VPP operations. A cooperative operation model has been proposed in \textbf{\cite{2023:Han}} to maximize the benefit of the VPP-CSs system by considering the conflicting interests of different stakeholders. In \textbf{\cite{2022b:Wang}}, a Stackelberg game model has been presented to establish an interaction between CSs and VPP with renewables, thermal generators, and energy storage units. A similar decentralized approach has been investigated in \textbf{\cite{2020:Fan}}, in which CSs are operated based on the price signals set by VPP. Nevertheless, in these studies, the CSs are limited to reacting passively to VPP price signals rather than actively engaging with the VPP operator, potentially compromising the effectiveness of EVs as energy demand and storage units. In order to overcome this issue, centralized control to coordinate the CS energy scheduling with other elements of VPP has been presented in \textbf{\cite{2021:Bin}}. However, previous studies regarding the incorporation of EVs into VPP operations overlooked the priorities of EV users as an individual unit. This includes factors such as the preferred arrival and departure times of EV owners at the CSs, as well as the maximum amount of energy that can be delivered to them at their departure time. Moreover, with the introduction of fast EV chargers, the utility grid may experience an increase in the peak-to-average ratio of energy demand during the charging period of EVs. Consequently, effectively addressing the scheduling problem and maximizing the utilization of EVs at CSs as storage hubs has proven challenging.

        The growing integration of unpredictable renewable sources and EVs within VPP introduces a continual risk associated with its engagement in the electricity market. In line with the imbalance market mechanism \textbf{\cite{2018:Pape}}, VPPs bear responsibility for their supply discrepancies, which must be reconciled within the imbalance mechanism through costly alternatives. Consequently, substantial fluctuations in DERs can jeopardize the VPP's viability. Hence, it is crucial to devise a practical and effective strategy for VPPs to mitigate this risk. Historically, traditional power plants have been employed to act as the operational reserve, yet this approach demands substantial investment and exacerbates carbon emission concerns. Energy storage can potentially mitigate discrepancies in DER generation \textbf{\cite{2020:Pearre}}; however, its feasibility is hindered by the current substantial upfront investment costs. Presently, leveraging smart technologies, demand response (DR) emerges as a cost-efficient strategy to address the aforementioned challenges \textbf{\cite{2017:Song}}. From the perspective of the VPP, DR can be seen as a virtual operational reserve, aiding in offsetting deviations in DER generation. Generally, the DR program consists of two main categories: price-based DR and incentive-based DR. Price-based DR lacks dispatchability, resulting in reduced flexibility. Moreover, there is often a lack of consumer understanding regarding their electricity usage and billing, which prompts inquiries into whether participation in price-based DR programs is motivated by perceived savings rather than tangible savings \textbf{\cite{2018:Lee}}. Conversely, incentive-based DR offers the ability to dispatch resources and significantly greater flexibility in assisting the VPP in acquiring the necessary DR resources \textbf{\cite{2018:Eissa}}. Hence, consumers are more inclined to engage in incentive-based DR programs than those based solely on pricing. Encouraging consumers to enroll in such incentive-based DR programs can enhance the stability of the VPP \textbf{\cite{2015:Ashot}}.

        Furthermore, it is imperative to account for the uncertainties associated with renewable energy sources, electricity demand, and pricing when devising optimal scheduling strategies for VPPs. Various methodologies, such as robust, possibilistic, and stochastic optimization, have been examined in the literature to analyze how uncertain variables affect the results of VPP scheduling problems \textbf{\cite{2019:Yu}}. Robust optimization is particularly applicable in scenarios where it is challenging to ascertain the distribution of uncertain parameters or when historical data is scarce in terms of both quantity and quality. Implementing robust optimization demands a clear understanding of uncertainty variables, encompassing their scale and extent, thereby presenting challenges in its application across specific power system problems \textbf{\cite{2019:Ju,2018:Koraki}}. Alternatively, possibilistic \textbf{\cite{2016:Fan}} and stochastic \textbf{\cite{2022:Riaz}} approaches quantify uncertain parameters by generating scenarios based on a membership function and probability density function (PDF), respectively, which are derived from historical data. Among these approaches, stochastic methods are comparatively simpler for power system issues. Stochastic techniques can additionally be categorized into analytical, numerical, and approximate methodologies. The analytical methods may sacrifice precision in calculating PDF of the uncertain input variable by employing simplifications such as linearizing system models and Gaussian distributions \textbf{\cite{2020:Wang,2017:Batzelis}}. Monte Carlo simulation is a widely employed and precise numerical method, particularly valued for its effectiveness in analyzing complex models with multiple random variables. This approach entails the generation of numerous random samples to represent uncertainties, thereby constructing a substantial dataset for assessing the PDFs of output parameters \textbf{\cite{2014:Amin}}. However, a drawback of numerical methods lies in the significant computational time needed to attain convergence. Nevertheless, approximate methods such as the point estimation method (PEM) provide a beneficial balance between precision and computational speed \textbf{\cite{2016:Aien}}. This work explicitly adopts Hong's (2m+1) PEM \textbf{\cite{2016:Ali}} that utilizes deterministic approaches to tackle probabilistic concerns and alleviate obstacles arising from incomplete knowledge of probability distributions linked with random variables.

        While extensive literature exists on the VPP scheduling problem, there has been limited attention given to the challenges arising from the involvement of multiple stakeholders in VPP operations. Furthermore, as previously mentioned, there is a necessity to implement a priority-based energy scheduling strategy within the VPP scheduling framework, taking into account the preferences of EV owners and other energy consumers regarding their energy demand or load curtailment requirements.
        Following the identification of research gaps within the VPP scheduling problem, this study introduces a stochastic multi-objective energy scheduling problem modeled for the direct-controlled VPP. The main contributions of this work are listed below:
        \begin{enumerate}
            \item A multi-objective problem is formulated in this work that accounts for the stochastic behavior of DR participants and emphasizes the conflicting objectives of various stakeholders within a VPP framework.
            \item This work presents a charge scheduling problem for EVs operating in CSs that prioritizes EV users based on their charging needs in the pre-scheduled operating time.
            \item An incentive-based DR program is proposed, which adopts a uniform incentivization approach to compensate controllable load owners for their inconvenience and EV users for their battery degradation.
        \end{enumerate}

    \section{Mathematical model of VPP}

\begin{figure}[!t]
            \centering
            \includegraphics[width=0.95\columnwidth]{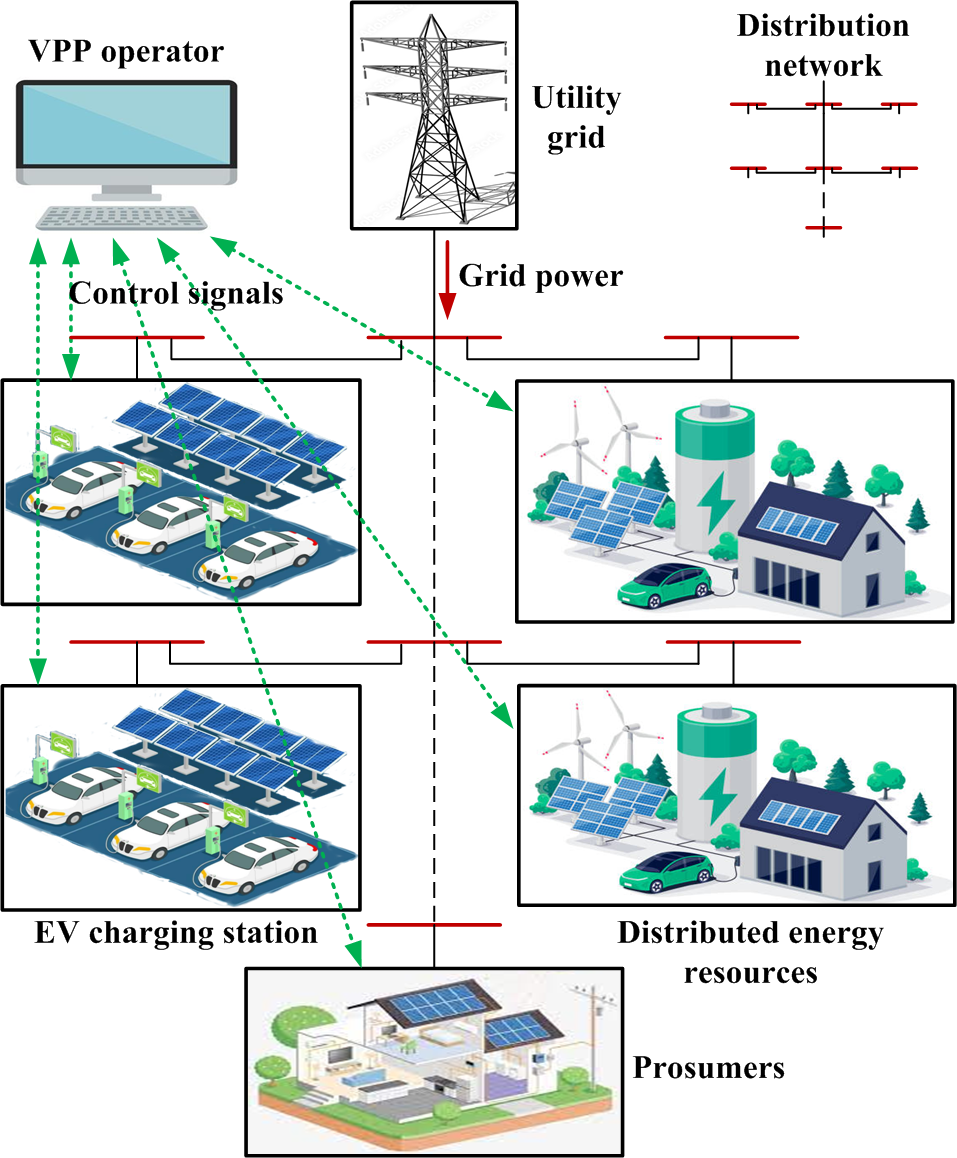}
            \vspace{-5pt}
            \caption{Direct-controlled virtual power plant.}
            \label{F:VPP}
        \end{figure}

        This section provides the mathematical model of solar-powered generating stations, EVs operating at CSs, batteries at BSSs, and controllable loads, that a VPP operator virtually aggregates through a direct-controlled approach. A schematic model of direct-controlled VPP, based on \textbf{\cite{2018:Thomas}}, is shown in \textbf{Fig. \ref{F:VPP}}. 
        
        \subsection{Solar-powered generating station}

            The power output of solar-powered generating stations at each operating interval predominantly relies on the solar irradiance level and ambient temperature at the site, which can be expressed mathematically as \textbf{\cite{2023:Harsh}}
            \begin{subequations}
                \label{E:Solar_Power}
                \begin{equation}
                    P_{s}^{n,t} = \ceil*{{P_{s1}^{n}}/{P_{s0}}} \times {FF} \times {V_s} \times {I_s}, \ \forall \ n\in\mathcal{S}^{N}; \ t\in\mathcal{S}^{T} \tag{\ref{E:Solar_Power}}
                \end{equation}
                where
                \begin{align}
                    FF= & \left({V_{MPP}{I_{MPP}}}\right)/\left({V_{OC}{I_{SC}}}\right) \\
                    V_s= & V_{OC} - k_v {T_c} \\
                    I_s= & s^t \left(I_{SC} + k_i \left(T_c-25\right)\right)\ \\
                    T_c= & T_a + s^t \left({T_{o}-20}\right)/{0.8}
                \end{align}
            \end{subequations} 

        \subsection{Electric vehicle charging station}
        \label{SS:EVCS}

            EVs can be modeled as battery units when it is connected to the grid through CS for their charging operation. It can be represented using its state of charge (SoC), which is mathematically calculated as \textbf{\cite{2023:Harsh}}
            \begin{multline}
                SoC_v^{n,t} = SoC_v^{n,t-1} + \frac{\left({\eta_{G2V}}\right)^{z_v^{n,t}}}{ \left({\eta_{V2G}}\right)^{1-z_v^{n,t}}} \left( \frac{P_v^{n,t} \Delta t}{E_v^{n}} \right), \\ \forall  \ v\in\mathcal{S}_V^n; \ n\in\mathcal{S}^{N}; \ t\in\mathcal{S}^{T}
            \end{multline}
            It is important to note here that $P_v^{n,t}$ is regarded as positive during G2V operation and negative during V2G operation of EVs.

            An EV, when connected to CS, can function as an energy storage unit that can operate in different operational modes: idle, uncoordinated G2V, and coordinated G2V/V2G mode. During idle mode, no power is exchanged between the EV and the CS, a condition that occurs when the EV is either fully charged or not connected to the CS. An uncoordinated mode represents the normal charging operation of an EV wherein it is charged at its rated power until it reaches its maximum SoC level. This operational method could potentially escalate peak energy demand with an increased number of EVs connected at CS. The coordinated G2V/V2G mode implements a controlled charging operation, aiming to alleviate grid burden by either lowering the aggregated energy demand of EVs at CS or by utilizing V2G operation to supply energy.

            This study suggests a prioritization mechanism for the CSs, shown in \textbf{Table \ref{T:mode}}, to control the transition between various modes of EV operation. 
            \begin{table}[!t]
                \centering
                \caption{Operational mode selection mechanism for EV}
                \vspace{-5pt}
                \label{T:mode}
                \begin{tabular}{|c|c|c|c|c|}
                    \hline
                     \multirow{3}{*}{$\rho_v^{n,t}$} & $\geq 1$ & \multicolumn{3}{c|}{Uncoordinated G2V mode}  \\
                     \cline{2-5}
                     & \multirow{2}{*}{$< 1$} & \multirow{2}{*}{$p_v^{n,t}$} & $1$ & Coordinated G2V mode \\
                     \cline{4-5}
                     & & & $0$ & Coordinated V2G mode \\
                     \hline
                \end{tabular}
            \end{table}
            The proposed mechanism depends on two factors: the priority factor ($\rho_v^{n,t}$) and the pricing factor ($p_v^{n,t}$). A priority factor is derived from three key determinants: the SoC of the EV at the beginning of the interval, the necessary charging duration ($\overline{t_{v}^n}$) to reach the maximum SoC, and the departure time of the EV from the CS. The proposed priority factor is mathematically modeled as
            \begin{subequations}
                \label{E:Priority_factor}
                \begin{equation}
                    \rho_v^{n,t}=\left(\overline{SoC}-SoC_v^{n,t}\right)^{\left(\left(t_{dep,v}^n-t\right)-\overline{t_{v}^n}\right)/T} \tag{\ref{E:Priority_factor}}
                \end{equation}
                \begin{equation}
                \hspace{-40pt}\text{where} \ \ \ \ \   \overline{t_{v}^n}=\left(\overline{SoC}-SoC_v^{n,t}\right){E_v^n}/{\left(\eta_{G2V} \overline{P_{v}^n} \right)}
                \end{equation}
            \end{subequations} 
            As the priority factor of an EV increases from zero to unity, the necessity for charging the EV intensifies. A priority factor exceeding one indicates that the EV lacks sufficient parking time at CS to reach its maximum SoC and, therefore, cannot undergo coordinated charging. Thus, \textbf{(\ref{E:Priority_factor})} explicitly determines the preferences of each EV owner in terms of their energy demand during the pre-scheduled parking duration.

            Further, a binary pricing factor is incorporated into the mode selection mechanism to account for electricity prices. This factor prohibits EV charging during peak price periods, thereby minimizing EV charging expenses. \textbf{Algorithm \ref{A:Pricing_factor}} outlines the methodology to determine the binary pricing factor for an EV during each operational interval.
            \begin{algorithm}[!t]
            \caption{Pricing factor computational algorithm for EV}
            \label{A:Pricing_factor}
                \begin{algorithmic}[1]
                   \STATE Initialize $p_v^{n,t}=0$.
                   \STATE Initialize $\mathcal{T}= \{i \ | \ \forall \ i\in[t,t_{dep,v}^n]\}$.
                   \STATE Initialize $\mathcal{Y}=\{\gamma^i \ | \ \forall \ i\in[t,t_{dep,v}^n]\}$.
                   \FOR{$i\in[1,t_{dep,v}^n-t]$}
                        \FOR{$j\in[i+1,t_{dep,v}^n-t+1]$}
                            \IF{$\mathcal{Y}_i\geq\mathcal{Y}_j$}
                                \STATE Exchange the elements at the $i^{th}$ index of $\mathcal{Y}$ and $\mathcal{T}$ with the elements at the $j^{th}$ index.
                            \ENDIF
                        \ENDFOR
                   \ENDFOR
                   \FOR{$i\in[1,\overline{t_{v}^n}]$}
                        \IF{$\mathcal{T}_i=t$}
                            \STATE Update $p_v^{n,t}=1$
                        \ENDIF
                    \ENDFOR
                \end{algorithmic} 
            \end{algorithm}

            By leveraging the different operational modes inherent in EVs, the CSs, acting as an aggregator of EVs, connected to designated nodes within the distribution network can engage in DR programs by modifying their energy consumption pattern to reduce the dependency on the utility grid. In exchange, the CSs may receive incentives from the VPP operator in the form of discounted electricity prices, structured as
            \begin{equation}
                cost_{CS}^n=\sum\limits_{t\in\mathcal{S}^T} \gamma^t \left( \sum\limits_{v\in\mathcal{S}_V^n} P_{v}^{n,t} \Delta t\right), \ \forall \ n \in \mathcal{S}^N
            \end{equation}

            However, the V2G operation of EVs will result in battery degradation that can be calculated in the form of battery degradation cost as \textbf{\cite{2015:Fard}}
            \begin{subequations}
                \label{E:bat_deg}
                \begin{multline}
                    cost_{deg,v}^{n}=\sum\limits_{t\in\mathcal{S^T}} \left(\left(DoD_{v}^{n,t} \right)^{1-b_d} - \left( DoD_{v}^{n,t-1}\right)^{1-b_d} \right) \\ \left(1-z_{v}^{n,t}\right) {c_b E_{v}^{n}}/{a_d}, \ \forall  \ v \in \mathcal{S}_V^n; \ n \in \mathcal{S}^N \tag{\ref{E:bat_deg}}
                \end{multline}
                \begin{equation}
                    \hspace{-120pt}\text{where} \ \ \ \ DoD_{v}^{n,t}=1-SoC_{v}^{n,t}
                \end{equation}
            \end{subequations}

        \subsection{Battery swapping station}

            In this work, a BSS has been incorporated alongside the solar-powered generating station to store surplus energy and facilitate its use during night hours. The batteries in BSS can be represented mathematically in terms of its SoC as
            \begin{multline}
                SoC_b^{n,t} = SoC_b^{n,t-1} + \frac{\left({\eta_{G2B}}\right)^{z_b^{n,t}}}{ \left({\eta_{B2G}}\right)^{1-z_b^{n,t}}} \left( \frac{P_b^{n,t} \Delta t}{E_b^{n}} \right), \\ \forall  \ b\in\mathcal{S}_B^n; \ n\in\mathcal{S}^{N}; \ t\in\mathcal{S}^{T}
            \end{multline}
            Here, $P_b^{n,t}$ is regarded as positive during G2B operation and negative during B2G operation.
            
            The batteries within the BSS can utilize the operational mode selection mechanism detailed in \textbf{Section \ref{SS:EVCS}} with minor adjustments. Given that the BSS primarily functions to store surplus energy without drawing power from the grid for charging, pricing considerations are irrelevant in the mode selection mechanism for the battery. Instead, the mechanism should incorporate solar irradiance levels to facilitate G2B mode during day hours and B2G mode during night hours, as shown in \textbf{Table \ref{T:modeBattery}}. The solar irradiance levels can determine the operating intervals when the solar is available for the scheduling operation and thus, makes it an important factor in determining the intervals for G2B and B2G mode operation.
            \begin{table}[!t]
                \centering
                \caption{Operational mode selection mechanism for battery}
                \vspace{-5pt}
                \label{T:modeBattery}
                \begin{tabular}{|c|c|c|c|c|}
                    \hline
                     \multirow{3}{*}{$\rho_b^{n,t}$} & $\geq 1$ & \multicolumn{3}{c|}{Uncoordinated G2B mode}  \\
                     \cline{2-5}
                     & \multirow{2}{*}{$< 1$} & \multirow{2}{*}{$s^t$} & $\geq \sigma$ & Coordinated G2B mode \\
                     \cline{4-5}
                     & & & $< \sigma$ & Coordinated B2G mode \\
                     \hline
                \end{tabular}
            \end{table}

            Moreover, the BSS offers an alternative charging solution for EV users who opt out of participating in the DR program due to their limited parking durations. So, the priority factor of the battery will depend on the information of the EV registered for battery swapping and is calculated as
            \begin{subequations}
                \label{E:Priority_factor_BSS}
                \begin{equation}
                    \rho_b^{n,t}=\left(\overline{SoC}-SoC_b^{n,t}\right)^{\left(\left(t_{arr,b}^n-t\right)-\overline{t_{b}^n}\right)/T} \tag{\ref{E:Priority_factor_BSS}}
                \end{equation}
                \begin{equation}
                    \hspace{-50pt}\text{where} \ \ \ \ \ \overline{t_{b}^n}=\left(\overline{SoC}-SoC_b^{n,t}\right){E_b^n}/{\left(\eta_{G2B} \overline{P_{b}^n} \right)}
                \end{equation}
            \end{subequations} 
            An important point to note here is that the SoC of the battery, after the swapping operation, will transition to a different level, typically lower, based on the SoC of the EV's battery registered for the swapping operation. Also, this work assumes that BSS is available for swapping operations in only daytime.
            
        \subsection{Controllable loads}

            Typically, energy consumers opt to engage in DR programs voluntarily, aiming to obtain benefits such as lower electricity expenses or incentives. These consumers can be analyzed based on their energy procurement costs from the utility grid, which can be expressed as
            \begin{equation}
                cost_{CL}^{n}=\sum\limits_{t\in\mathcal{S}^T} \alpha^t \left(P_{0}^{n,t}-x^{n,t}\right) \Delta t, \ \forall \ n\in\mathcal{S}^L
            \end{equation}
            Here, $x^{n,t}$ is positive for the load curtailment and negative for load increment. Furthermore, to provide flexibility to consumers in terms of DR participation time, this work assumes that the consumers with controllable loads need to have advance registration of their preferred time slot [$t_{start}^n,t_{stop}^n$] for load curtailments.

            To lower the electricity expenses for these consumers, the VPP operator coordinates the usage of their controllable loads by adjusting them to intervals when electricity prices are low. This determination of low-priced intervals relies on a binary pricing factor determined by \textbf{Algorithm \ref{A:Pricing_factor_curt}}. Thereafter, intervals, where the pricing factor is unity, are designated for load curtailment, while the remaining intervals are utilized for shifting the curtailed loads.  
            \begin{algorithm}[!t]
            \caption{Pricing factor computational algorithm for controllable loads}
            \label{A:Pricing_factor_curt}
                \begin{algorithmic}[1]
                   \STATE Initialize $p_{CL}^{n,t}=0$; and $T_0=t_{stop}^n-t_{start}^n+1$.
                   \STATE Initialize $\mathcal{T}= \{i \ | \ \forall \ i\in[t_{start}^n,t_{stop}^n]\}$.
                   \STATE Initialize $\mathcal{Y}=\{\alpha^i \ | \ \forall \ i\in[t_{start}^n,t_{stop}^n]\}$.
                   \FOR{$i\in[1,T_0-1]$}
                        \FOR{$j\in[i+1,T_0]$}
                            \IF{$\mathcal{Y}_i\leq\mathcal{Y}_j$}
                                \STATE Exchange the elements at the $i^{th}$ index of $\mathcal{Y}$ and $\mathcal{T}$ with the elements at the $j^{th}$ index.
                            \ENDIF
                        \ENDFOR
                   \ENDFOR
                   \FOR{$i\in[1,T_0]$}
                        \IF{$\mathcal{T}_i=t$}
                            \STATE Update $p_{CL}^{n,t}=1$; and $T_0=T_0-1$.
                        \ENDIF
                    \ENDFOR
                \end{algorithmic} 
            \end{algorithm}

            Further, the shifting of controllable loads may result in discomfort to consumers involved in the operation, which can be mathematically expressed in the form of discomfort cost as \textbf{\cite{2016:Yu}}
            \begin{equation}
                cost_{dis}^{n}= \sum\limits_{t\in\mathcal{S}^T}\left(e^{\beta^n\left(p_{CL}^{n,t}{x^{n,t}}/{P_{0}^{n,t}}\right)}-1\right)
            \end{equation}
            
    \section{Multi-objective problem formulation}

        This section introduces the mathematical formulations of the proposed multi-objective VPP scheduling problem in a day-ahead market. For a better understanding, the deterministic formulation, ignoring the uncertainties in the scheduling problem, is provided in the first subsection. However, in the day-ahead market, uncertainties prevail regarding generation, load, and electricity prices, rendering them unknown. A stochastic model, using Hong's $(2m+1)$ PEM \textbf{\cite{2016:Ali}}, is formulated in this work to consider these uncertainties, which is discussed in a later subsection. 

        \subsection{Deterministic problem formulation}

            The proposed work considers VPP scheduling for a day-ahead market from the perspective of different stakeholders, i.e., EVs, CSs, controllable loads, VPP operators, and DS operators, involved in the VPP operation. 

            From the VPP operator's viewpoint, they aim to optimize profitability by meeting the demand of CSs using power from solar generating stations. Any surplus solar energy can be stored in batteries located at BSS and subsequently supplied to the DS at a reduced electricity rate. The profit function of the VPP operator is mathematically modeled as
            \begin{subequations}
            \label{E:F1}
                \begin{equation}
                    F_1 = \begin{cases} 
                        \sum\limits_{n\in\mathcal{S}^N} \sum\limits_{t\in\mathcal{S}^T}  \gamma^t P_{inj}^{n,t} \Delta t, & \text{if $z^{n,t}=0$} \\ \noalign{\vskip5pt}
                        \sum\limits_{n\in\mathcal{S}^N} \sum\limits_{t\in\mathcal{S}^T} \alpha^t P_{inj}^{n,t} \Delta t , & \text{if $z^{n,t}=1$} 
                    \end{cases} \tag{\ref{E:F1}}
                \end{equation}
                \begin{equation}
                    \hspace{-20pt}\text{where} \ \ \ \ P_{inj}^{n,t}=\left(P_{s}^{n,t}-\sum\limits_{v\in\mathcal{S}^n_V}P_{v}^{n,t}-\sum\limits_{b\in\mathcal{S}_B^n}P_{b}^{n,t}\right)
                \end{equation}
                
            \end{subequations}

            However, the VPP also aims to establish a balanced approach in determining the appropriate scale for solar power generation facilities that aligns with their installation expenses. In mathematical terms, it can be articulated by considering daily capital investments alongside capital recovery factors and maintenance expenses, as expressed in \textbf{(\ref{E:F2})}.
            \begin{subequations}
                \label{E:F2}
                \begin{equation}
                    F_2 = \sum\limits_{n\in\mathcal{S}^N}\left( c_i c_p + c_0  \right)P_{s1}^n/365 \tag{\ref{E:F2}}
                \end{equation}
                \begin{equation}
                    \hspace{-28pt}\text{where capital recovery factor, }c_i=\frac{r(1+r)^{L^t}}{(1+r)^{L^t}-1}
                \end{equation}
            \end{subequations}

            Furthermore, the CSs and consumers with controllable loads, who are the key participants in the DR program, perceive the VPP scheduling problem as a chance to minimize their daily energy expenses. As elaborated in the preceding section, this study integrates incentivized electricity rates for CSs, while consumers with controllable loads gain advantages by adjusting their energy usage to periods of lower prices. Mathematically, the functional objective of DR participants is formulated as
            \begin{equation}
                F_3= \sum\limits_{n\in\mathcal{S}^{N}}cost_{CS}^n + \sum\limits_{n\in\mathcal{S}^L}cost_{CL}^n
            \end{equation}

            In the proposed DR program, the utilization of V2G technology could lead to unforeseen expenses related to battery degradation of EVs. Furthermore, consumers with controllable loads may experience inconvenience or discomfort as a result of load shifting, necessitating efforts to minimize such impacts. A mathematical representation of undesirable conditions arises due to the DR program being modeled in the form of a cost function as
            \begin{equation}
                F_4= \sum\limits_{n\in\mathcal{S}^{N}}\sum\limits_{v\in\mathcal{S}_V^{n}}cost_{deg,v}^n + \sum\limits_{n\in\mathcal{S}^L}cost_{dis}^n
            \end{equation}

            The DS operator seeks to diminish reliance on the utility grid, a goal that can be quantified in terms of grid power consumption expenses as
            \begin{equation}
                F_5= \sum\limits_{t\in\mathcal{S}^{T}} \alpha^t P_{grid}^t \Delta t
            \end{equation}

            Given the diverse and often conflicting goals of the various stakeholders engaged in the operation of the VPP, the proposed multi-objective optimization problem is formulated as
            \begin{equation}
            \label{E:obj}
                Minimize \ \{-F_1,F_2,F_3,F_4,F_5\}
            \end{equation}

            However, due to the inherent conflict within defined objective functions, simultaneous optimization of the proposed VPP scheduling problem is unattainable. This study employs the utopia-tracking method introduced in \textbf{\cite{2012:Victor}} to address the conflicting nature of multi-objective functions by consolidating them into a normalized mono-objective function. This strategy facilitates attaining a trade-off Pareto optimal solution that closely aligns with utopia points. The utopia-tracking methodology involves normalizing the individual objective function by considering optimal and suboptimal solutions as
            \begin{equation}
                f_i = \begin{cases} 
                    \frac{\left(F_i-\underline{F_i}\right)}{\left(\overline{F_i}-\underline{F_i}\right)}, & \text{if $F_i$ is to be minimized} \\ \noalign{\vskip5pt}
                    \frac{\left(\overline{F_i}-F_i\right)}{\left(\overline{F_i}-\underline{F_i}\right)}, & \text{if $F_i$ is to be maximized}
                \end{cases} 
            \end{equation}

            Based on the normalized objectives obtained using the utopia-tracking approach, the deterministic form of the multi-objective function outlined in \textbf{(\ref{E:obj})} is reformulated as
            \begin{subequations}
                \label{E:objective}
                \begin{equation}
                    Minimize \ \mathcal{F}=\sum\limits_{i=1}^5 w_i f_i \tag{\ref{E:objective}}
                \end{equation}
                subject to
                \begin{equation}
                \label{E:constraint1}
                    \overline{P_s} \leq P_s^{n} \leq \underline{P_s}, \ \ \ \ \forall \ n\in\mathcal{S}^N 
                \end{equation}
                \begin{equation}
                \label{E:constraint2}
                    \overline{P_v^{n}} \leq P_v^{n,t} \leq \underline{P_v^n}, \ \ \ \ \forall \ n\in\mathcal{S}^N; \ v\in\mathcal{S}_V^n; \ t\in\mathcal{S}^T 
                \end{equation}
                \begin{equation}
                \label{E:constraint3}
                    \overline{P_b^{n}} \leq P_b^{n,t} \leq \underline{P_b^n}, \ \ \ \ \forall \ n\in\mathcal{S}^N; \ b\in\mathcal{S}_B^n; \ t\in\mathcal{S}^T 
                \end{equation}
                \begin{equation}
                \label{E:constraint4}
                    \mu_1 \alpha^t \leq \gamma^{t} \leq \mu_2 \alpha^t, \ \ \ \ \forall \ t\in\mathcal{S}^T 
                \end{equation}
                \begin{equation}
                \label{E:constraint5}
                    \overline{x^{n,t}} \leq x^{n,t} \leq \underline{x^{n,t}}, \ \ \ \ \forall \ n\in\mathcal{S}^{L}; \ t\in\mathcal{S}^T  
                \end{equation}
                \begin{equation}
                \label{E:constraint6}
                    \overline{SoC} \leq SoC_v^{n,t} \leq \underline{SoC}, \ \ \ \forall \ n\in\mathcal{S}^N; \ v\in\mathcal{S}_V^n; \ t\in\mathcal{S}^T  
                \end{equation}
                \begin{equation}
                \label{E:constraint7}
                    \overline{SoC} \leq SoC_b^{n,t} \leq \underline{SoC}, \ \ \ \forall \ n\in\mathcal{S}^N; \ b\in\mathcal{S}_B^n; \ t\in\mathcal{S}^T  
                \end{equation}
                \begin{equation}
                \label{E:constraint8}
                    SoC_v^{n,t_{dep,v}^n}=\overline{SoC}, \ \ \ \ \forall \ n\in\mathcal{S}^N; \ v\in\mathcal{S}_V^n
                \end{equation}
                \begin{equation}
                \label{E:constraint9}
                    SoC_b^{n,t_{arr,b}^n}=\overline{SoC}, \ \ \ \ \forall \ n\in\mathcal{S}^N; \ b\in\mathcal{S}_B^n
                \end{equation}
                \begin{equation}
                \label{E:constraint10}
                    \sum\limits_{t\in\mathcal{S}^T} x^{n,t} = 0, \ \ \ \ \forall \ n\in\mathcal{S}^L
                \end{equation}
                \begin{equation}
                \label{E:constraint11}
                    \sum\limits_{t\in\mathcal{S}^T} p_{CL}^{n,t} x^{n,t} \leq \lambda_1 \sum\limits_{t\in\mathcal{S}^T} P_0^{n,t}, \ \ \ \ \forall \ n\in\mathcal{S}^L
                \end{equation}
                \begin{equation}
                \label{E:constraint12}
                    \frac{\sum\limits_{t\in\mathcal{S}^T} \alpha^t x^{n_1,t}}{\sum\limits_{t\in\mathcal{S}^T} \alpha^t x^{n_2,t}} = \frac{cost_{dis}^{n_1}}{cost_{dis}^{n_2}}, \ \forall \ \beta^{n_1}=\beta^{n_2}; \ n_1,n_2\in\mathcal{S}^L 
                \end{equation}
                \begin{equation}
                \label{E:constraint13}
                    P_s^{n,t} \! \geq \! \lambda_2 \! \! \left( \sum\limits_{v\in\mathcal{S}_V^n} \! P_v^{n,t} \! + \! \sum\limits_{b\in\mathcal{S}_B^n} \! P_b^{n,t} \!  \right) \! \!, \ \forall \! \ n\in\mathcal{S}^N; \! \ t\in\mathcal{S}^T 
                \end{equation}
                \begin{equation}
                \label{E:constraint14}
                    \sum\limits_{n\in\mathcal{S}^N} \sum\limits_{t\in\mathcal{S}^T}  \left(1-z^{n,t}\right)\gamma^t P_{inj}^{n,t} \Delta t  \leq  DB
                \end{equation}
            \end{subequations} 
            
            The operational constraints outlined in \textbf{(\ref{E:constraint1})}-\textbf{(\ref{E:constraint5})} delineate the boundary thresholds for various decision variables of the proposed optimization problem. The maximum capacity of a solar station is determined by the extent of the sun-exposed area accessible to the operator, as outlined in \textbf{(\ref{E:constraint1})}. The boundary limit of batteries within EVs and BSS are contingent upon their manufacturer-specified power ratings and is integrated into the problem using \textbf{(\ref{E:constraint2})} and \textbf{(\ref{E:constraint3})}, respectively. Furthermore, the incentivized electricity rate offered by the VPP operator must consistently remain lower than the prevailing market electricity rate by the margins outlined in \textbf{(\ref{E:constraint4})}. In this work, the value of $\mu_1$ and $\mu_2$ is fixed to 0.6 and 0.9, respectively. The power curtailment limits for consumers with controllable loads have been incorporated in the work using \textbf{(\ref{E:constraint5})}. Considering the work presented in \textbf{\cite{2016:Alireza}}, this work assumes that sixty percent of loads in each operating interval are controllable. Further, to restrict the proposed DR program from forming another peak during an off-peak period, this work formulates the curtailment limits as
            \begin{equation}
                \overline{x^{n,t}}=0.6 P_0^{n,t}; \ \ \underline{x^{n,t}}=P_0^{n,t}-max\{P_0^{n,:}\} 
            \end{equation}
            
            To restrict the batteries from over-charging and over-discharging, constraints \textbf{(\ref{E:constraint6})} and \textbf{(\ref{E:constraint7})} ensure that the SoC of batteries within EVs and BSS are within the prescribed limits. Additionally, the operator must guarantee that EVs departing from the CS at their designated time and EVs enlisted for battery swapping receive batteries with the maximum SoC level. This assurance is achieved using \textbf{(\ref{E:constraint8})} and \textbf{(\ref{E:constraint9})}, respectively.

            Equation \textbf{(\ref{E:constraint10})} guarantees that consumers will not experience any reduction in load when calculated on a day scale. Furthermore, in this study, \textbf{(\ref{E:constraint11})}  is implemented to ensure that the overall load shifts in a day remain within tolerable levels of discomfort for consumers. This work assumes that all consumers share an equal level of acceptable discomfort, and thus, $\lambda_1$ is set to 0.2. However, to maintain uniformity among consumers with controllable loads, the VPP operator must ensure that the benefits provided to consumers align proportionately with any inconvenience resulting from load control implemented by the VPP. This task is managed using \textbf{(\ref{E:constraint12})}.

            Further, when determining the ideal capacity for a solar-powered generating station, the VPP operator must address concerns regarding under-sizing, which could arise when minimizing the objective function $F_2$. This condition is managed by introducing \textbf{(\ref{E:constraint13})} as an operating constraint that ensures the fulfillment of certain factors of CSs and BSSs demand during the daytime. Thus, in this work, $\lambda_2$ is assigned a value of 0.6 during daytime and 0 during nighttime. Finally, considering the limited daily budget of the DS operator, constraint \textbf{(\ref{E:constraint14})} restricts the financial transaction from the DS operator to the VPP operator.

        \subsection{Stochastic problem formulation}

            In the stochastic scenario, uncertain input variables are characterized by probabilistic information derived from their mean, standard deviation, and PDF.  This work focuses on the uncertainties in the energy demand of DR participants, i.e., EVs and controllable loads, whose PDFs are considered to be a Gaussian distribution function, represented as
            \begin{equation}
                f_{u}=\frac{1}{\sigma_{u} \sqrt{2\pi}} \exp \left(-\frac{\left(u-\mu_{u}\right)^2}{2\sigma_{u}^2}\right), \ \forall \ u\in\mathcal{S}^U
            \end{equation}

            Hong's $(2m+1)$ PEM uses this information to analytically define the different concentrations (or scenarios) in the form of locations, with their respective weights, of the uncertain input variables as \textbf{\cite{2016:Ali}}
            \begin{subequations}
                \label{E:PEM}
                \begin{equation}
                    \begin{aligned}
                    p_{u,k}= & \mu_{u}+\xi_{u,k} \sigma_{u},\ \forall\ u\in\mathcal{S}^U;\ k=\{1,2\} \\
                    \omega_{u,k}= & \left\{
                            \begin{array}{c l}
                                 \frac{\left(-1\right)^{3-k}}{\xi_{u,k}.\left(\xi_{u,1}-\xi_{u,2}\right)}, & \ \  k=1,2 \\ \noalign{\vskip5pt}
                                 \frac{1}{m}-\frac{1}{\lambda_{u,4}-\left(\lambda_{u,3}\right)^2}, & \ \ k=3 
                            \end{array}
                            \right. 
                    \end{aligned}
                    \tag{\ref{E:PEM}}
                \end{equation}
                where 
                \begin{align}
                    \xi_{u,k} =   & \left\{ \! \!  \!  \! 
                        \begin{array}{c l} 
                             \frac{\lambda_{u,3}}{2} \! + \! \left(\! -1\! \right)^{3\! -\! k} \sqrt{ \! \lambda_{u,4}\! -\! \frac{3}{4}\!  \left(\! \lambda_{u,3}\! \right)^2},\!  & \! \!  k\! =\! 1,2  \\ \noalign{\vskip5pt}
                             0,\! \! & \!  k=3 
                        \end{array}   
                        \right. \\
                    \lambda_{u,j}= & \frac{M_j\left(u\right)}{\left(\sigma_{u}\right)^j}, \ \forall\ j=\{1,2,3,4\} \\
                    M_j\left(u\right)= & \int_{-\infty}^{\infty}{\left(u-\mu_{u}\right)^j f_{u} {du}}, \ \forall\ j=\{1,2,3,4\}
                \end{align}
            \end{subequations}

            Using the various locations derived from Hong's $(2m+1)$ PEM, the deterministic problem outlined in \textbf{(\ref{E:objective})} is evaluated for $(2m+1)$ iterations, each involving distinct values of the uncertain variables, as depicted by
            \begin{subequations}
                \label{E:Final_Objective}
                \begin{equation}
                    Minimize \ \mathcal{F}_{obj} = \sum\limits_u \sum\limits_k \omega_{u,k}  \mathcal{F}(p_{u,k},\mathcal{S}^{'})  \tag{\ref{E:Final_Objective}}
                \end{equation}
            subject to \textbf{(\ref{E:constraint1})}-\textbf{(\ref{E:constraint14})}. \\
            \end{subequations}
            Here, $\mathcal{S}^{'}= \{\mu_i | i\in\mathcal{S}^{U},i\neq u\}$. The proposed stochastic formulation aims to minimize the mean value of the multi-objective VPP scheduling problem. However, it can also be used to minimize the other probabilistic information of the scheduling problem.
            
    \section{Results and validations}

        This section provides an overview of the input data for the proposed multi-objective VPP scheduling problem. Subsequent subsections discuss simulation results acquired from MATLAB and the real-time digital simulator (RTDS).

        \subsection{System data}

            The proposed problem is solved on a 12.66 kV, 33-node DS integrated with solar-powered generating stations, CSs, and BSSs at different node locations. However, to reduce DC-AC conversion losses, this work assumes the integration of solar stations, CSs, and BSSs at node numbers 8,15,21,23, and 30 as a single unit, thus, forming solar-powered battery-integrated EV CSs at each node location. Further, it is presumed that each CS is outfitted with ten distinct models of EV chargers, with the capacity to accommodate 100 EVs daily. Detailed technical specifications for the ten EV models under consideration can be found in \textbf{\cite{2023:Harsh}}. The uncertainty in the arrival and departure schedules of EVs at the CSs is addressed by utilizing the data provided in \textbf{Table \ref{T:probabilistic}}. Additionally, to ensure the availability of the appropriate battery for each type of EV during the swapping operation at the BSS, ten batteries with matching specifications to those of EV batteries are stocked at the BSS. Furthermore, to determine the optimal size for solar-powered generating stations, this study takes into account the technical specifications and cost information of solar modules provided in \textbf{\cite{2023:Vignesh}}.

            \begin{table}[!t]
                \centering
                \caption{Probabilistic information of EVs and controllable loads}
                \vspace{-5pt}
                \label{T:probabilistic}
                \begin{tabular}{|c | c |c c c|}
                \hline
                \multicolumn{2}{|c|}{} & \thead{Mean} & \thead{Standard\\ deviation} & \thead{Min/Max}  \\
                \hline
                \multirow{3}{*}{{EV}} & \thead{Arrival time} & \thead{08:00} & \thead{03:00} & \thead{01:00/20:00} \\
                & \thead{Departure time} & \thead{17:00} & \thead{03:00} & \thead{11:00/24:00} \\
                & \thead{SoC at arrival} & \thead{0.5} & \thead{0.25} & \thead{0.3/0.9} \\
                \hline
                \multirow{2}{*}{{\thead{Controllable\\ loads}}} & DR start time & \thead{08:00} & \thead{03:00} & \thead{01:00/20:00} \\
                & \thead{DR stop time} & \thead{17:00} & \thead{03:00} & \thead{11:00/24:00} \\
                \hline
                \end{tabular}
            \end{table}

            This work also assumes the participation of all consumers of the DS in the DR program, focusing on various consumer categories such as residential, commercial, and industrial, with more details available in \textbf{\cite{2017:Raoof}}. A standard deviation of 2\% is incorporated into the hourly energy demands of consumers to account for their variability. Additionally, the start and stop times for their DR operations are determined based on probabilistic data provided in \textbf{Table \ref{T:probabilistic}}. Other important parameters considered in this study are available in \textbf{\cite{2023:Harsh}}.

        \subsection{Simulation results}

            The proposed VPP scheduling optimization problem is addressed utilizing an improved Jaya algorithm, the algorithm of which is available in \textbf{\cite{2023:Harsh}}. The simulation work is performed using MATLAB R2023b on a desktop equipped with an Intel Core i9-14900K 3.20 GHz CPU and 64GB of memory.

\begin{figure}[t]
                \centering
                \includegraphics[width=\columnwidth]{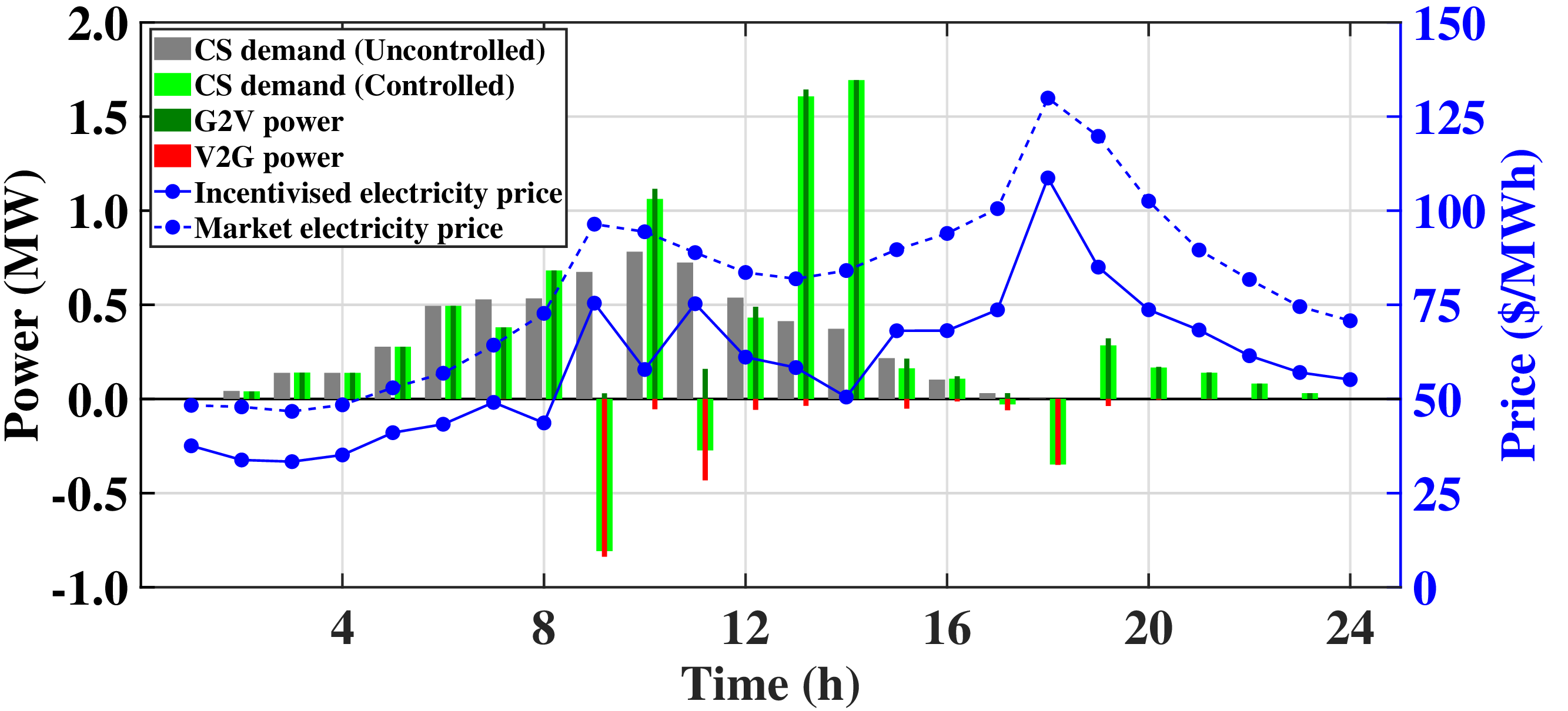}
                \vspace{-10pt}
                \caption{Hourly CS energy demand and incentivized electricity price for a day.}
                \label{F:CS_demand}
            \end{figure}

\begin{figure}[t]
                \centering
                \includegraphics[width=\columnwidth]{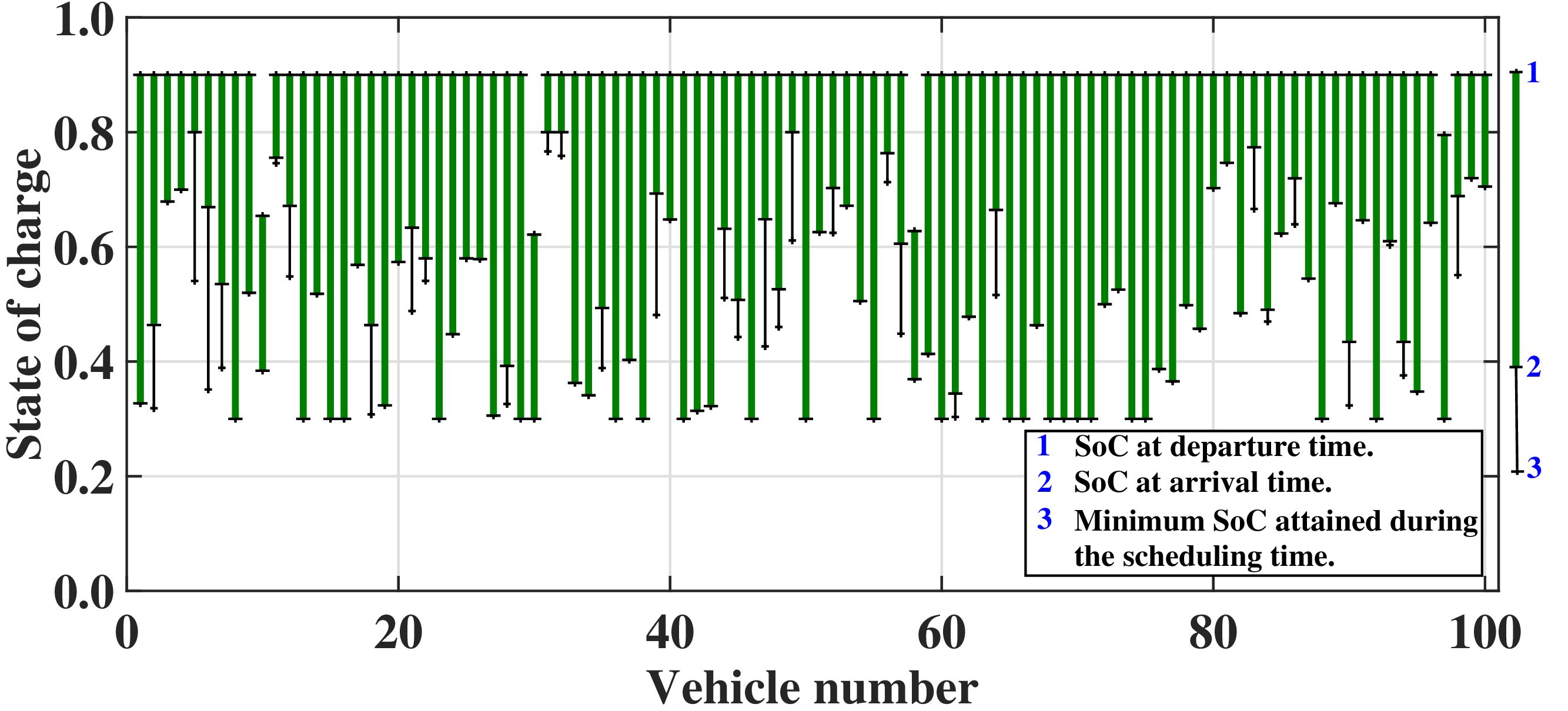}
                \vspace{-10pt}
                \caption{Variation in SOC of EVs during their stay at the CS.}
                \label{F:EV_SoC}
            \end{figure}

            The primary aim of the proposed work is to establish an efficient approach for controlling the charging operation of EVs at the CS, taking into account the individual priorities of each EV user. \textbf{Figure \ref{F:CS_demand}} illustrates the daily distribution of hourly energy demand from CS, showcasing the efficient utilization of EVs as storage units during controlled G2V/V2G operations in contrast to their uncontrolled operations. The VPP operator offers an incentivized electricity price to encourage the participation of CSs in the DR program, which is depicted in \textbf{Fig. \ref{F:CS_demand}} and is notably lower than the prevailing market electricity price. The proposed incentive strategy ensures that EVs conduct G2V operations during periods of low pricing and V2G operations during high pricing periods. Additionally, the strategy guarantees that EVs achieve the maximum SoC feasible upon departure, as evidenced by the SoC depicted in the candlestick chart presented in \textbf{Fig. \ref{F:EV_SoC}}, illustrating the status of 100 EVs at the CS integrated into node number 8.

\begin{figure}[t]
                \centering
                \includegraphics[width=\columnwidth]{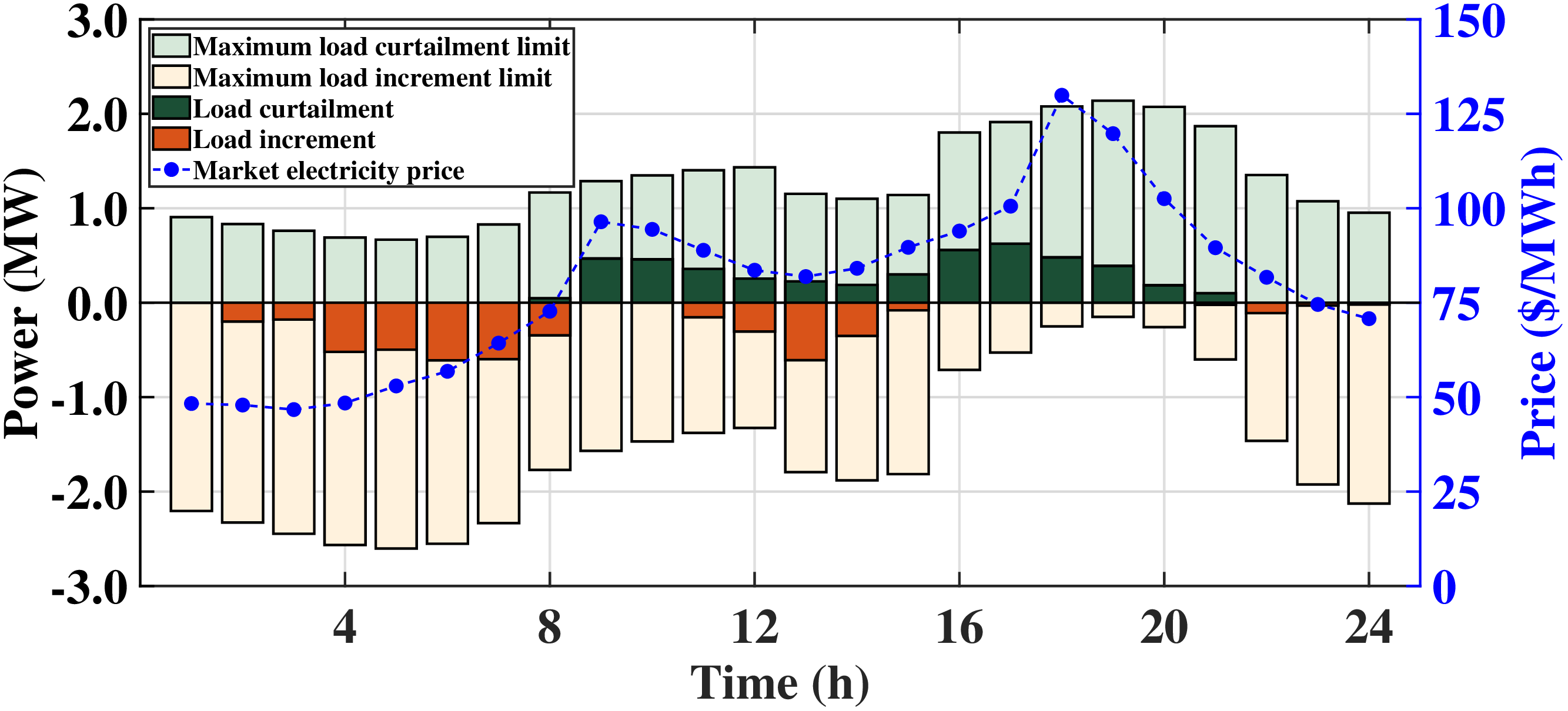}
                \vspace{-10pt}
                \caption{Hourly load curtailments and increments for the consumers with controllable loads.}
                \label{F:CL_demand}
            \end{figure}

            Consumers possessing controllable loads, integrated at various nodes of the DS, actively engage in the DR program by adjusting their load operation schedules to coincide with low-price periods and, thus, are rewarded by the reduced electricity bill for a day operation. \textbf{Figure \ref{F:CL_demand}} illustrates the overall energy adjustment within each operational interval, highlighting instances of load reduction during periods of peak pricing. It also verifies the successful establishment of boundary limits for load shifting during any operational interval, aiming to prevent the occurrence of another peak in the off-peak period.

\begin{figure}[t]
                \centering
                \includegraphics[width=\columnwidth]{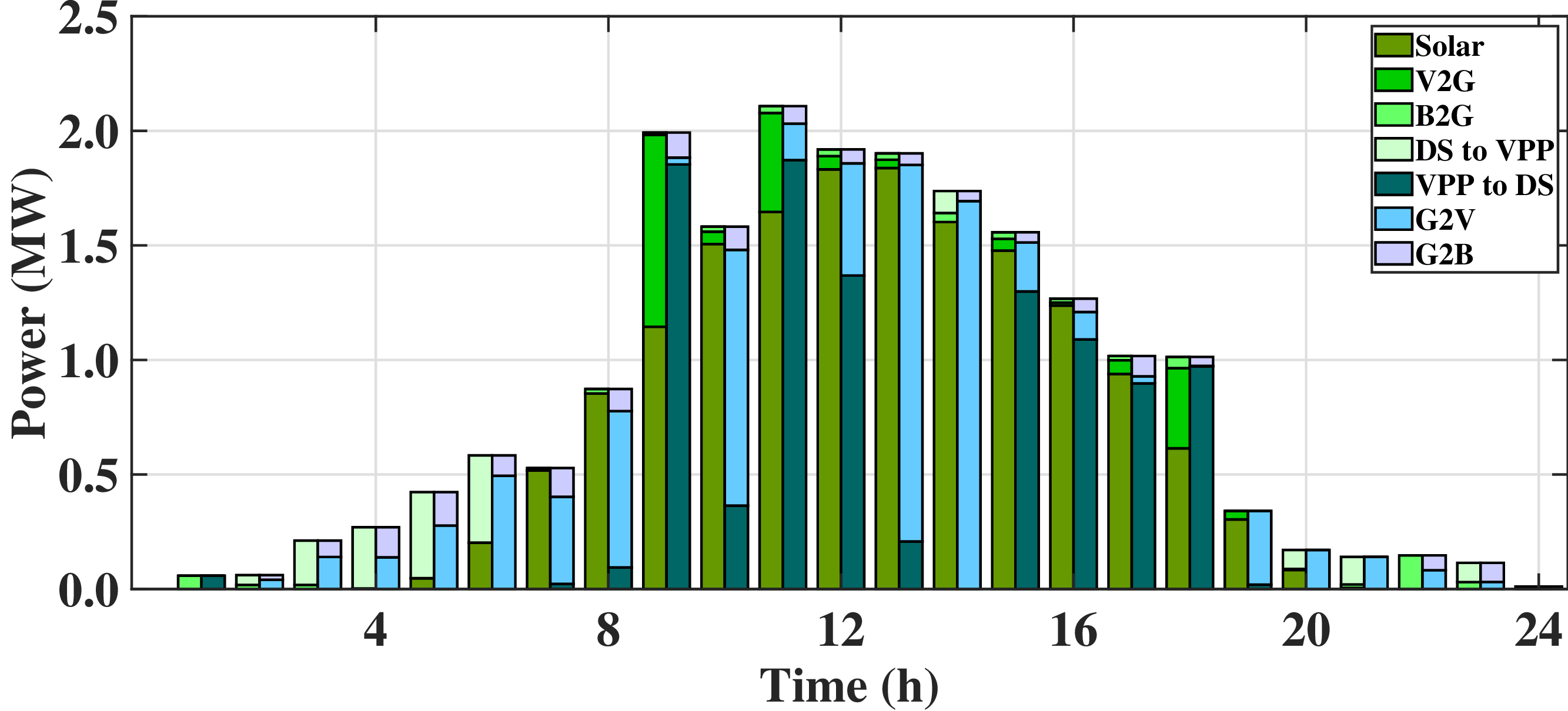}
                \vspace{-10pt}
                \caption{Hourly power exchange between solar, EVs, batteries, and DS as a source and load within the VPP framework.}
                \label{F:VPP_demand}
            \end{figure}

            Further, the proposed VPP scheduling approach recommends the optimal size for the solar-powered generating station to maximize the satisfaction of energy requirements for CSs and BSSs through the renewable source, illustrated in \textbf{Fig. \ref{F:VPP_demand}}. \textbf{Figure \ref{F:VPP_demand}} also confirms the power demand and supply balance in the VPP framework achieved by taking advantage of V2G and B2G operation of EVs and batteries as a power source at each operating interval. The surplus solar energy is stored within the batteries at the BSS to be utilized by CSs during night hours. Additionally, following the satisfaction of the energy demands of CSs and BSSs, the suggested scheduling approach further enhances the profitability of the VPP by supplying electricity to DS at incentivized pricing. The quantitative measure of the total energy supplied by the VPP to the DS during a day, along with the corresponding profit generated from this energy transaction is shown in \textbf{Table \ref{T:Comparison}}.

            \begin{table}[t]
                \centering
                \caption{Comparison of the proposed approach with the uncontrolled approach in a day-ahead operation}
                \vspace{-5pt}
                \label{T:Comparison}
                \footnotesize
                \begin{tabular}{|l|c|c|c|}
                    \hline
                    & \thead{Uncontrolled\\Approach} & \thead{Proposed\\Approach} & \thead{\%\\reduction} \\
                    \hline
                    \thead[l]{Energy purchased from the\\ grid (MWh)} & \thead{\textcolor{blue}{67.97}\\ \textcolor{red}{(0.84)}} & \thead{\textcolor{blue}{53.70}\\ \textcolor{red}{(2.35)}} & \thead{20.99} \\
                    \thead[l]{Energy delivered by the VPP\\ to the DS (MWh)} & - & \thead{\textcolor{blue}{10.43}\\ \textcolor{red}{(0.58)}} & - \\
                    \thead[l]{Energy lost in the DS (MWh)} & \thead{\textcolor{blue}{2.5314}\\ \textcolor{red}{(0.0517)}} & \thead{\textcolor{blue}{1.9637}\\ \textcolor{red}{(0.1067)}} & \thead{22.43} \\
                    \hline
                    \thead[l]{Cost of energy purchased\\ from the grid (\$)} & \thead{\textcolor{blue}{5614.14}\\ \textcolor{red}{(80.44)}} & \thead{\textcolor{blue}{4116.38}\\ \textcolor{red}{(167.15)}} & \thead{26.68}  \\ 
                    \thead[l]{VPP profit made by deliver-\\ing energy to the DS (\$)} & - & \thead{\textcolor{blue}{583.72}\\ \textcolor{red}{(109.37)}} & - \\
                    \thead[l]{Cost of charging EVs (\$)} & \thead{\textcolor{blue}{411.63}\\ \textcolor{red}{(76.47)}} & \thead{\textcolor{blue}{269.94}\\ \textcolor{red}{(68.06)}} & \thead{34.42} \\
                    \thead[l]{Electricity cost of consumers\\ with controllable loads (\$)} & \thead{\textcolor{blue}{4949.36}\\ \textcolor{red}{(7.62)}} & \thead{\textcolor{blue}{4786.96}\\ \textcolor{red}{(22.10)}} & \thead{3.28} \\
                    
                    \hline

                    \multicolumn{4}{l}{\thead[l]{$\#$ \textcolor{blue}{Blue} and \textcolor{red}{Red} colour represent the mean and standard deviation,\\ respectively.}} \\
                \end{tabular}
            \end{table}

            \textbf{Table \ref{T:Comparison}} also demonstrates improvements in the energy reliance of DS on the grid and the energy dissipated in the distribution network. Additionally, it presents advancements in the operational costs for various stakeholders. Furthermore, this study effectively addresses the impact of uncertain energy demands from controllable loads and CSs, which is presented in terms of mean and standard deviations in energy and cost shown in \textbf{Table \ref{T:Comparison}}.

        \subsection{Result validation}

            A software-in-the-loop simulation utilizing MATLAB, RSCAD, and RTDS is conducted to validate the accuracy and efficacy of the proposed VPP scheduling methodology. It is important to note here that this study employs the conventional Newton-Raphson power flow technique for determining the voltages at each node within the DS. The Newton-Raphson method addresses the linearized power flow model derived from approximating the non-linear power flow equations. However, this approximation can potentially result in overestimation or underestimation of node voltage values, thereby posing a risk to the accuracy of the results. Thus, in this work, the VPP network comprising DS, CS, BSS, and solar-powered generating stations is first modeled in RSCAD, followed by real-time EMT simulation executed in RTDS, where non-linear differential equations associated with the VPP model are solved at each time-step (typically within $50 \mu s$) so that this simulation is free from the errors due to the linear approximations.

\begin{figure}[!t]
                \centering
                \includegraphics[width=0.85\columnwidth]{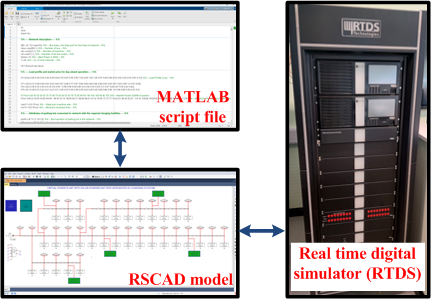}
                \vspace{-5pt}
                \caption{Software-in-loop setup with MATLAB, RSCAD and RTDS.}
                \label{F:RTDS_RSCAD}
            \end{figure}

            The MATLAB and RSCAD-RTDS co-simulation platform, as shown in \textbf{Fig. \ref{F:RTDS_RSCAD}}, is developed using the logical and physical ports of the Gigabit transceiver network communication (GTNETx2) card, which is installed in the RTDS rack. The GTNETx2 card provides support for the TCP/IP Socket protocol, allowing for bidirectional communication between MATLAB and RSCAD-RTDS. This capability enables the transmission and reception of stored real-time network variable data from RSCAD to MATLAB, as well as from MATLAB to RSCAD, at every $1s$ interval.

\begin{figure}[!t]
                \centering
                \includegraphics[width=\columnwidth]{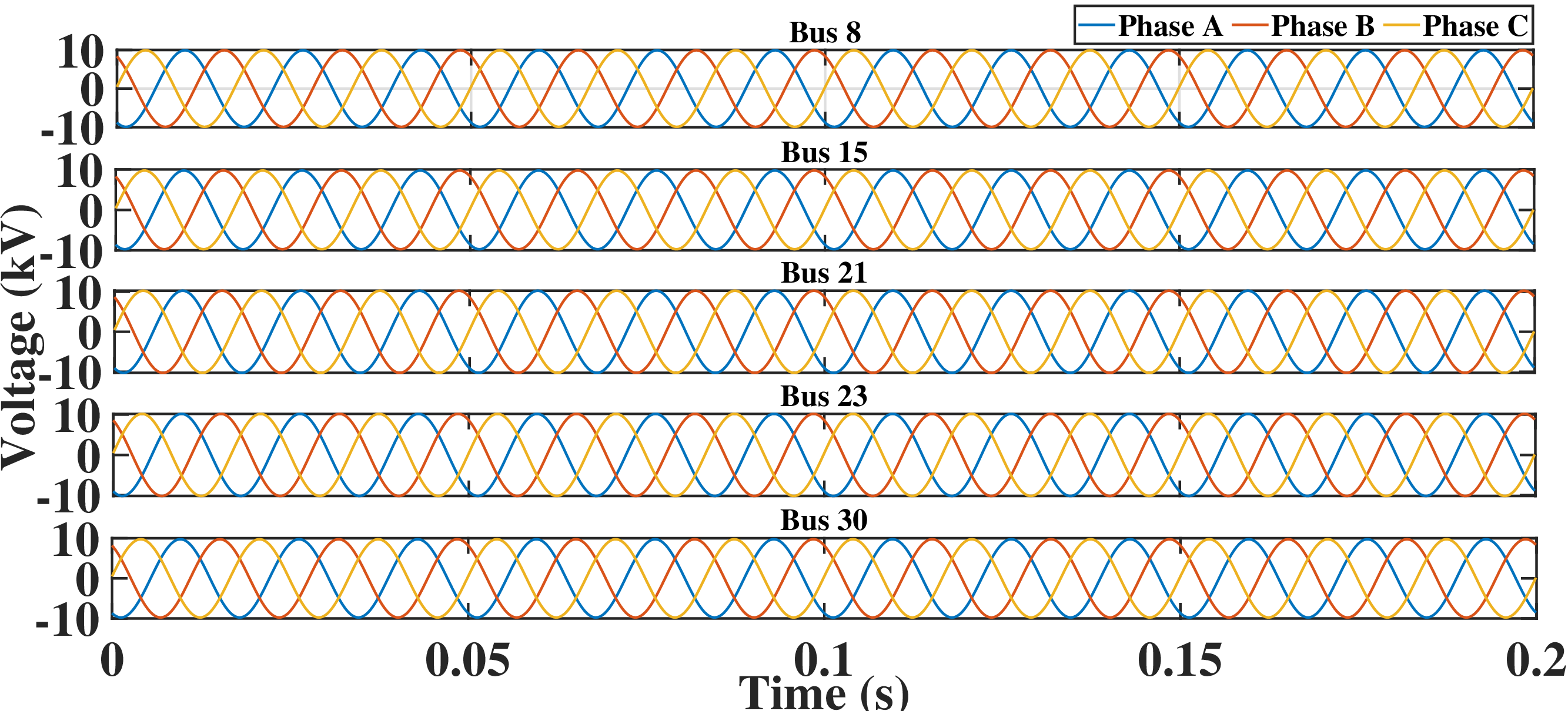}
                \vspace{-10pt}
                \caption{Node voltage of CSs at the Hour 11.}
                \label{F:CS_voltage}
            \end{figure}

\begin{figure}[!t]
                \centering
                \subfloat[]{
                    \includegraphics[width=\columnwidth]{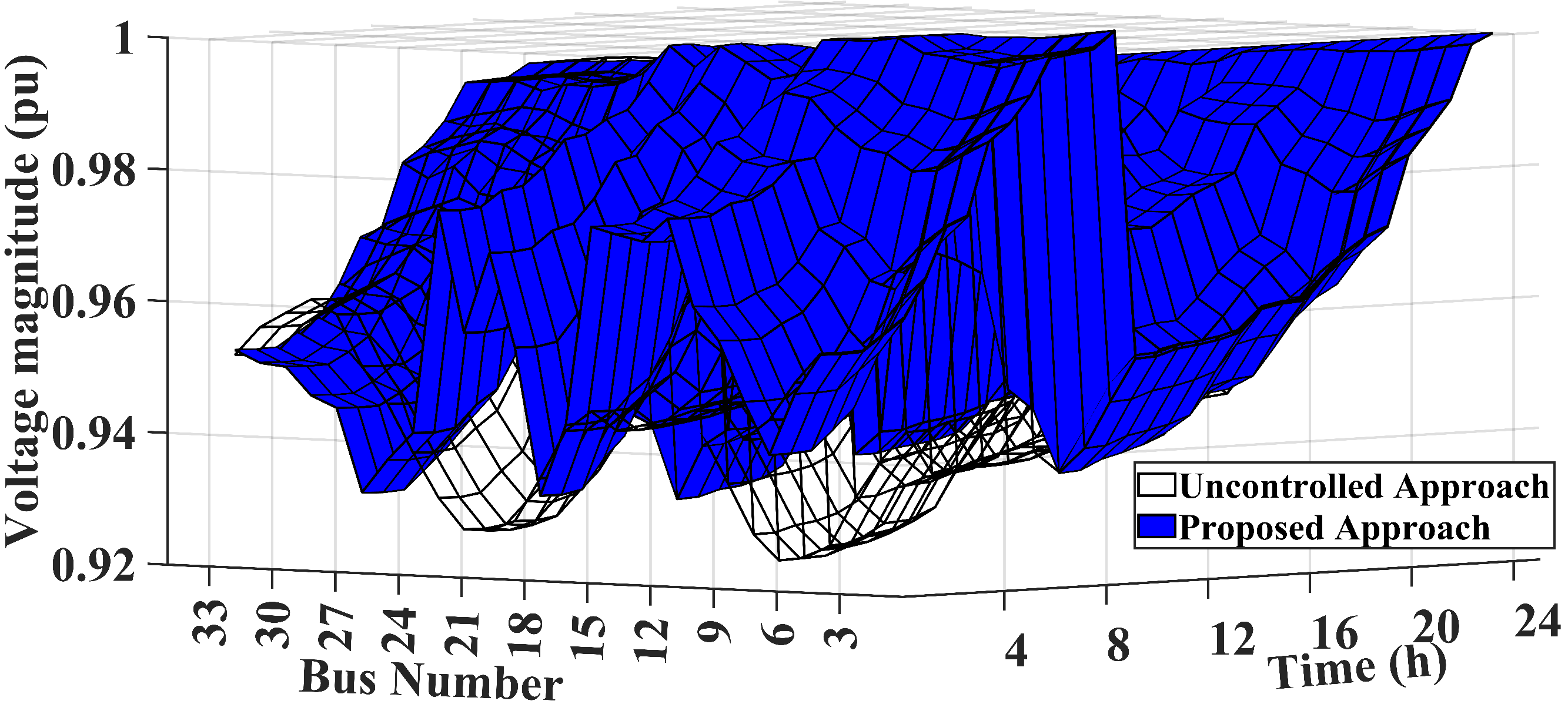}
                }
                \\
                \subfloat[]{
                    \includegraphics[width=\columnwidth]{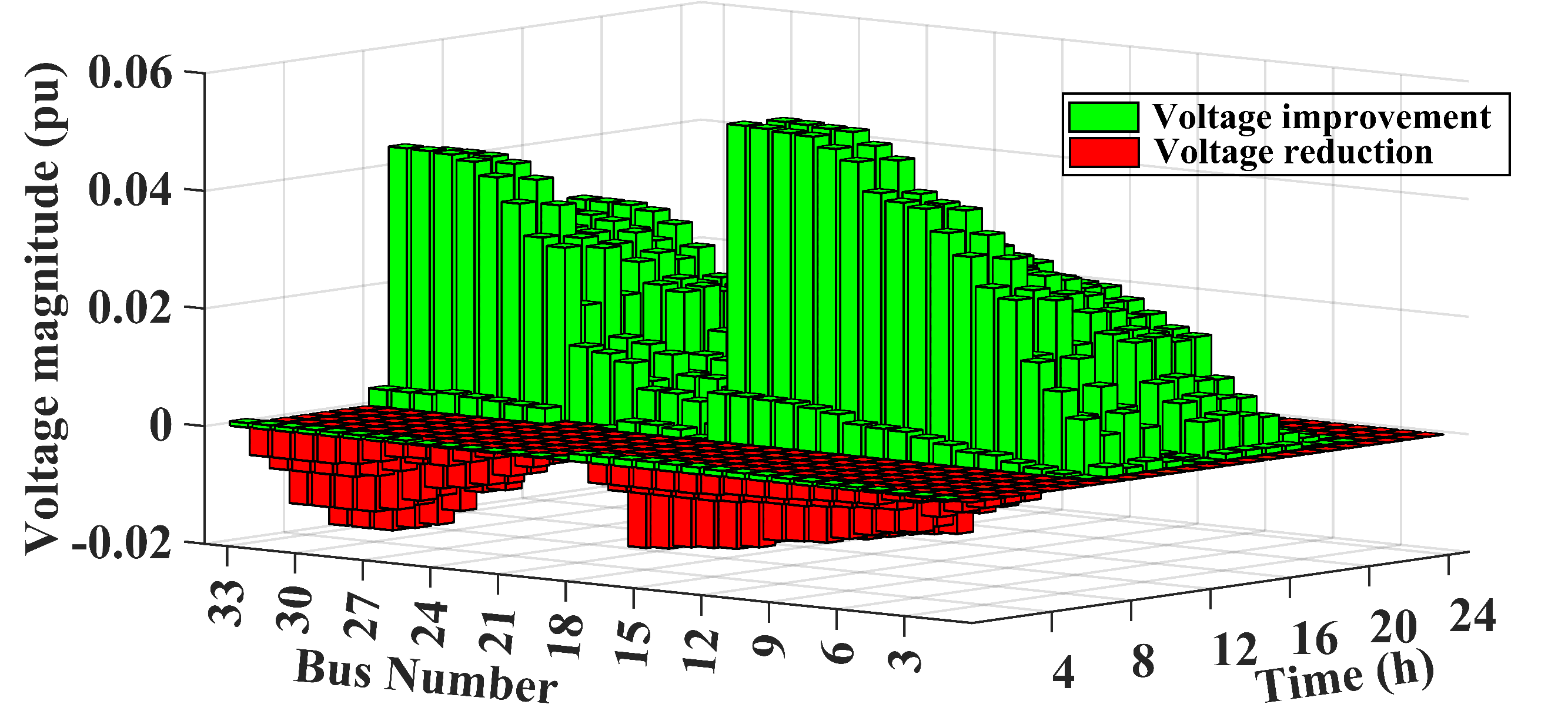}
                }
                \vspace{-10pt}
                \caption{(a) Hourly voltage profile, \& (b) Voltage improvement obtained in the proposed approach compared to the uncontrolled approach.}
                \label{F:voltage_profile}
            \end{figure}

            The software-in-loop simulation validates that the DS operates without transients during each operational interval. The nodes most susceptible to vulnerabilities, which are integrated with CSs and serve as common points for energy exchange between VPP and DS, have been chosen to showcase the transient-free voltage waveform depicted in \textbf{Fig. \ref{F:CS_voltage}}. Further, from the \textbf{Fig. \ref{F:VPP_demand}}, it is concluded that the maximum energy exchange between the VPP and DS occurs during the eleventh hour of the day, indicating this period as a potential operational interval with transiencies. However, \textbf{Fig. \ref{F:CS_voltage}} validates the steady operation of DS at the eleventh hour of the day. Further, the improvement in the voltage profile of all the nodes in DS, particularly during day hours, is showcased in \textbf{Fig. \ref{F:voltage_profile}}. The night hours experience some reduction in the voltages due to the load-shifting operation during the low-price period. It also confirms that the voltage limits have not been violated during the proposed operation.

    \section{Conclusion}

        A stochastic priority-ordered incentive-based DR program is presented in this work for the effective energy scheduling in the VPP comprised of solar-powered generating stations, BSSs, EV CSs, and consumers with controllable loads. The priority mechanism is suggested for the EVs at different CSs to effectively utilize the EVs as a storage unit in the VPP scheduling approach while ensuring the attainment of maximum SoC at their departure time. A multi-objective VPP scheduling problem is presented in this work to account for the conflicting objectives of different stakeholders and a utopia-tracking approach is used to find a near-optimal solution for each objective. Further, the incentives offered by the VPP operator at the end of scheduling lead to the reduction in the charging cost of EVs by 34.42\% and the electricity cost of consumers with controllable loads by 3.28\%. The simulation results verify that all the operational constraints are satisfied during the operation. Furthermore, the co-simulation using RSCAD-RTDS validates the steady-state operation of DS in each operating interval. Future work will focus on increasing the resiliency of the VPP operation during unexpected or undesirable weather conditions. 

    \section*{Acknowledgment} \addcontentsline{toc}{section}{Acknowledgment} 
        This work was supported by the Engineering and Physical Sciences Research Council [grant number EP/Y005376/1] – VPP-WARD Project (\href{https://www.vppward.com}{\textcolor{blue}{https://www.vppward.com}}).

    
    \bibliographystyle{IEEEtran}
    \bibliography{Reference.bib}

 \newcommand{\noop}[1]{}
\begin{thebibliography}{10}
\providecommand{\url}[1]{#1}
\csname url@samestyle\endcsname
\providecommand{\newblock}{\relax}
\providecommand{\bibinfo}[2]{#2}
\providecommand{\BIBentrySTDinterwordspacing}{\spaceskip=0pt\relax}
\providecommand{\BIBentryALTinterwordstretchfactor}{4}
\providecommand{\BIBentryALTinterwordspacing}{\spaceskip=\fontdimen2\font plus
\BIBentryALTinterwordstretchfactor\fontdimen3\font minus
  \fontdimen4\font\relax}
\providecommand{\BIBforeignlanguage}[2]{{%
\expandafter\ifx\csname l@#1\endcsname\relax
\typeout{** WARNING: IEEEtran.bst: No hyphenation pattern has been}%
\typeout{** loaded for the language `#1'. Using the pattern for}%
\typeout{** the default language instead.}%
\else
\language=\csname l@#1\endcsname
\fi
#2}}
\providecommand{\BIBdecl}{\relax}
\BIBdecl

\bibitem{2022:WEC}
\BIBentryALTinterwordspacing
``World energy trilemma index,'' World Energy Council, London, UK, Tech. Rep.,
  2022. [Online]. Available:
  \url{https://www.worldenergy.org/assets/downloads/World_Energy_Trilemma_Index_2022.pdf}
\BIBentrySTDinterwordspacing

\bibitem{2016:NERC}
\BIBentryALTinterwordspacing
``Distributed energy resources: Connection modeling and reliability
  considerations,'' North American Electric Reliability Corporation, Atlanta,
  USA, Tech. Rep., 2017. [Online]. Available:
  \url{https://www.nerc.com/comm/Other/essntlrlbltysrvcstskfrcDL/Distributed_Energy_Resources_Report.pdf}
\BIBentrySTDinterwordspacing

\bibitem{2010:Han}
D.-m. Han and J.-h. Lim, ``Smart home energy management system using {IEEE}
  802.15.4 and zigbee,'' \emph{IEEE Transactions on Consumer Electronics},
  vol.~56, no.~3, pp. 1403--1410, 2010.

\bibitem{2018:Zafar}
R.~Zafar, A.~Mahmood, S.~Razzaq, W.~Ali, U.~Naeem, and K.~Shehzad, ``Prosumer
  based energy management and sharing in smart grid,'' \emph{Renewable and
  Sustainable Energy Reviews}, vol.~82, pp. 1675--1684, 2018.

\bibitem{2020:Fernando}
F.~Lezama, J.~Soares, B.~Canizes, and Z.~Vale, ``Flexibility management model
  of home appliances to support {DSO} requests in smart grids,''
  \emph{Sustainable Cities and Society}, vol.~55, p. 102048, 2020.

\bibitem{2018:Helena}
H.~Gerard, E.~I. {Rivero Puente}, and D.~Six, ``Coordination between
  transmission and distribution system operators in the electricity sector: A
  conceptual framework,'' \emph{Utilities Policy}, vol.~50, pp. 40--48, 2018.

\bibitem{2016:Christakou}
K.~Christakou, ``A unified control strategy for active distribution networks
  via demand response and distributed energy storage systems,''
  \emph{Sustainable Energy, Grids and Networks}, vol.~6, pp. 1--6, 2016.

\bibitem{2024:Xie}
Y.~Xie, Y.~Zhang, W.-J. Lee, Z.~Lin, and Y.~A. Shamash, ``Virtual power plants
  for grid resilience: A concise overview of research and applications,''
  \emph{IEEE/CAA Journal of Automatica Sinica}, vol.~11, no.~2, pp. 329--343,
  2024.

\bibitem{2022:Shah}
\BIBentryALTinterwordspacing
J.~Shah, ``Introducing {VPP}ieces: Bite-sized blogs about virtual power
  plants,'' Loan Program Office, U.S. Department of Energy, Washington D.C.,
  U.S., Tech. Rep., 2022. [Online]. Available:
  \url{https://www.energy.gov/lpo/articles/introducing-vppieces-bite-sized-blogs-about-virtual-power-plants}
\BIBentrySTDinterwordspacing

\bibitem{2018:Thomas}
T.~Morstyn, N.~Farrell, S.~J. Darby, and M.~D. McCulloch, ``Using peer-to-peer
  energy-trading platforms to incentivize prosumers to form federated power
  plants,'' \emph{Nature Energy}, vol.~3, no.~2, pp. 94--101, 2018.

\bibitem{2023:Yi}
Y.~Kuang, X.~Wang, H.~Zhao, T.~Qian, N.~Li, J.~Wang, and X.~Wang, ``Model-free
  demand response scheduling strategy for virtual power plants considering risk
  attitude of consumers,'' \emph{CSEE Journal of Power and Energy Systems},
  vol.~9, no.~2, pp. 516--528, 2023.

\bibitem{2022:Huishi}
H.~Liang and J.~Ma, ``Data-driven resource planning for virtual power plant
  integrating demand response customer selection and storage,'' \emph{IEEE
  Transactions on Industrial Informatics}, vol.~18, no.~3, pp. 1833--1844,
  2022.

\bibitem{2020:Zhong}
W.~Zhong, M.~A.~A. Murad, M.~Liu, and F.~Milano, ``Impact of virtual power
  plants on power system short-term transient response,'' \emph{Electric Power
  Systems Research}, vol. 189, p. 106609, 2020.

\bibitem{2021:Sampath}
S.~K. Venkatachary, A.~Alagappan, and L.~J.~B. Andrews, ``Cybersecurity
  challenges in energy sector (virtual power plants) - can edge computing
  principles be applied to enhance security?'' \emph{Energy Informatics},
  vol.~4, no.~1, pp. 5--26, 2021.

\bibitem{2023:Deng}
J.~Deng and Q.~Guo, ``Decentralized energy management system of distributed
  energy resources as virtual power plant: Economic risk analysis via downside
  risk constraints technique,'' \emph{Computers \& Industrial Engineering},
  vol. 183, p. 109522, 2023.

\bibitem{2022:Wang}
X.~Wang, H.~Zhao, H.~Lu, Y.~Zhang, Y.~Wang, and J.~Wang, ``Decentralized
  coordinated operation model of {VPP} and {P2H} systems based on
  stochastic-bargaining game considering multiple uncertainties and carbon
  cost,'' \emph{Applied Energy}, vol. 312, p. 118750, 2022.

\bibitem{2012:Heussen}
K.~Heussen, S.~You, B.~Biegel, L.~H. Hansen, and K.~B. Andersen, ``Indirect
  control for demand side management - a conceptual introduction,'' in
  \emph{2012 3rd IEEE PES Innovative Smart Grid Technologies Europe}, 2012, pp.
  1--8.

\bibitem{2013:Dimitrios}
D.~Papadaskalopoulos, G.~Strbac, P.~Mancarella, M.~Aunedi, and V.~Stanojevic,
  ``Decentralized participation of flexible demand in electricity
  markets—{Part II}: Application with electric vehicles and heat pump
  systems,'' \emph{IEEE Transactions on Power Systems}, vol.~28, no.~4, pp.
  3667--3674, 2013.

\bibitem{2012:Mardavij}
M.~Roozbehani, M.~A. Dahleh, and S.~K. Mitter, ``Volatility of power grids
  under real-time pricing,'' \emph{IEEE Transactions on Power Systems},
  vol.~27, no.~4, pp. 1926--1940, 2012.

\bibitem{2016:Kostas}
K.~Margellos and S.~Oren, ``Capacity controlled demand side management: A
  stochastic pricing analysis,'' \emph{IEEE Transactions on Power Systems},
  vol.~31, no.~1, pp. 706--717, 2016.

\bibitem{2018:Matteo}
M.~Muratori, ``Impact of uncoordinated plug-in electric vehicle charging on
  residential power demand,'' \emph{Nature Energy}, vol.~3, no.~3, pp.
  193--201, 2018.

\bibitem{2023:Harsh}
P.~Harsh and D.~Das, ``A priority-ordered incentive-based smart charging
  strategy of electric vehicles to determine the optimal size of solar power
  plant at the charging stations,'' in \emph{2023 IEEE International Conference
  on Environment and Electrical Engineering and 2023 IEEE Industrial and
  Commercial Power Systems Europe}, 2023, pp. 1--6.

\bibitem{2021:Saleh}
S.~Sadeghi, H.~Jahangir, B.~Vatandoust, M.~A. Golkar, A.~Ahmadian, and
  A.~Elkamel, ``Optimal bidding strategy of a virtual power plant in day-ahead
  energy and frequency regulation markets: A deep learning-based approach,''
  \emph{International Journal of Electrical Power \& Energy Systems}, vol. 127,
  p. 106646, 2021.

\bibitem{2020:Yang}
D.~Yang, S.~He, M.~Wang, and H.~Pandžić, ``Bidding strategy for virtual power
  plant considering the large-scale integrations of electric vehicles,''
  \emph{IEEE Transactions on Industry Applications}, vol.~56, no.~5, pp.
  5890--5900, 2020.

\bibitem{2020:Sheidaei}
F.~Sheidaei and A.~Ahmarinejad, ``Multi-stage stochastic framework for energy
  management of virtual power plants considering electric vehicles and demand
  response programs,'' \emph{International Journal of Electrical Power \&
  Energy Systems}, vol. 120, p. 106047, 2020.

\bibitem{2021:Niti}
\BIBentryALTinterwordspacing
``Handbook of electric vehicle charging infrastructure implementation,'' NITI
  Aayog, India, Tech. Rep., 2021. [Online]. Available:
  \url{http://www.niti.gov.in/sites/default/files/2021-08/HandbookforEVChargingInfrastructureImplementation081221.pdf}
\BIBentrySTDinterwordspacing

\bibitem{2023:Han}
H.~Wang, Y.~Jia, M.~Shi, C.~S. Lai, and K.~Li, ``A mutually beneficial
  operation framework for virtual power plants and electric vehicle charging
  stations,'' \emph{IEEE Transactions on Smart Grid}, vol.~14, no.~6, pp.
  4634--4648, 2023.

\bibitem{2022b:Wang}
J.~Wang, C.~Guo, C.~Yu, and Y.~Liang, ``Virtual power plant containing electric
  vehicles scheduling strategies based on deep reinforcement learning,''
  \emph{Electric Power Systems Research}, vol. 205, p. 107714, 2022.

\bibitem{2020:Fan}
S.~Fan, J.~Liu, Q.~Wu, M.~Cui, H.~Zhou, and G.~He, ``Optimal coordination of
  virtual power plant with photovoltaics and electric vehicles: A temporally
  coupled distributed online algorithm,'' \emph{Applied Energy}, vol. 277, p.
  115583, 2020.

\bibitem{2021:Bin}
B.~Zhou, K.~Zhang, K.~W. Chan, C.~Li, X.~Lu, S.~Bu, and X.~Gao, ``Optimal
  coordination of electric vehicles for virtual power plants with dynamic
  communication spectrum allocation,'' \emph{IEEE Transactions on Industrial
  Informatics}, vol.~17, no.~1, pp. 450--462, 2021.

\bibitem{2018:Pape}
C.~Pape, ``The impact of intraday markets on the market value of flexibility
  — decomposing effects on profile and the imbalance costs,'' \emph{Energy
  Economics}, vol.~76, pp. 186--201, 2018.

\bibitem{2020:Pearre}
N.~Pearre and L.~Swan, ``Combining wind, solar, and in-stream tidal electricity
  generation with energy storage using a load-perturbation control strategy,''
  \emph{Energy}, vol. 203, p. 117898, 2020.

\bibitem{2017:Song}
M.~Song and M.~Amelin, ``Purchase bidding strategy for a retailer with flexible
  demands in day-ahead electricity market,'' \emph{IEEE Transactions on Power
  Systems}, vol.~32, no.~3, pp. 1839--1850, 2017.

\bibitem{2018:Lee}
L.~V. White and N.~D. Sintov, ``Inaccurate consumer perceptions of monetary
  savings in a demand-side response programme predict programme acceptance,''
  \emph{Nature Energy}, vol.~3, no.~12, pp. 1101--1108, 2018.

\bibitem{2018:Eissa}
M.~Eissa, ``First time real time incentive demand response program in smart
  grid with “i-energy” management system with different resources,''
  \emph{Applied Energy}, vol. 212, pp. 607--621, 2018.

\bibitem{2015:Ashot}
A.~Mnatsakanyan and S.~W. Kennedy, ``A novel demand response model with an
  application for a virtual power plant,'' \emph{IEEE Transactions on Smart
  Grid}, vol.~6, no.~1, pp. 230--237, 2015.

\bibitem{2019:Yu}
S.~Yu, F.~Fang, Y.~Liu, and J.~Liu, ``Uncertainties of virtual power plant:
  Problems and countermeasures,'' \emph{Applied Energy}, vol. 239, pp.
  454--470, 2019.

\bibitem{2019:Ju}
L.~Ju, R.~Zhao, Q.~Tan, Y.~Lu, Q.~Tan, and W.~Wang, ``A multi-objective robust
  scheduling model and solution algorithm for a novel virtual power plant
  connected with power-to-gas and gas storage tank considering uncertainty and
  demand response,'' \emph{Applied Energy}, vol. 250, pp. 1336--1355, 2019.

\bibitem{2018:Koraki}
D.~Koraki and K.~Strunz, ``Wind and solar power integration in electricity
  markets and distribution networks through service-centric virtual power
  plants,'' \emph{IEEE Transactions on Power Systems}, vol.~33, no.~1, pp.
  473--485, 2018.

\bibitem{2016:Fan}
S.~Fan, Q.~Ai, and L.~Piao, ``Fuzzy day-ahead scheduling of virtual power plant
  with optimal confidence level,'' \emph{IET Generation, Transmission \&
  Distribution}, vol.~10, no.~1, pp. 205--212, 2016.

\bibitem{2022:Riaz}
M.~Riaz, S.~Ahmad, I.~Hussain, M.~Naeem, and L.~Mihet-Popa, ``Probabilistic
  optimization techniques in smart power system,'' \emph{Energies}, vol.~15,
  no.~3, 2022.

\bibitem{2020:Wang}
C.~Wang, C.~Liu, F.~Tang, D.~Liu, and Y.~Zhou, ``A scenario-based analytical
  method for probabilistic load flow analysis,'' \emph{Electric Power Systems
  Research}, vol. 181, p. 106193, 2020.

\bibitem{2017:Batzelis}
E.~I. Batzelis, G.~E. Kampitsis, and S.~A. Papathanassiou, ``Power reserves
  control for {PV} systems with real-time {MPP} estimation via curve fitting,''
  \emph{IEEE Transactions on Sustainable Energy}, vol.~8, no.~3, pp.
  1269--1280, 2017.

\bibitem{2014:Amin}
M.~A. Tajeddini, A.~Rahimi-Kian, and A.~Soroudi, ``Risk averse optimal
  operation of a virtual power plant using two stage stochastic programming,''
  \emph{Energy}, vol.~73, pp. 958--967, 2014.

\bibitem{2016:Aien}
M.~Aien, A.~Hajebrahimi, and M.~Fotuhi-Firuzabad, ``A comprehensive review on
  uncertainty modeling techniques in power system studies,'' \emph{Renewable
  and Sustainable Energy Reviews}, vol.~57, pp. 1077--1089, 2016.

\bibitem{2016:Ali}
A.~G. Zamani, A.~Zakariazadeh, and S.~Jadid, ``Day-ahead resource scheduling of
  a renewable energy based virtual power plant,'' \emph{Applied Energy}, vol.
  169, pp. 324--340, 2016.

\bibitem{2015:Fard}
A.~Kavousi-Fard, T.~Niknam, and M.~Fotuhi-Firuzabad, ``Stochastic
  reconfiguration and optimal coordination of {V2G} plug-in electric vehicles
  considering correlated wind power generation,'' \emph{IEEE Transactions on
  Sustainable Energy}, vol.~6, no.~3, pp. 822--830, 2015.

\bibitem{2016:Yu}
M.~Yu and S.~H. Hong, ``A real-time demand-response algorithm for smart grids:
  A stackelberg game approach,'' \emph{IEEE Transactions on Smart Grid},
  vol.~7, no.~2, pp. 879--888, 2016.

\bibitem{2012:Victor}
V.~M. Zavala and A.~Flores-Tlacuahuac, ``Stability of multiobjective predictive
  control: A utopia-tracking approach,'' \emph{Automatica}, vol.~48, no.~10,
  pp. 2627--2632, 2012.

\bibitem{2016:Alireza}
A.~Soroudi, P.~Siano, and A.~Keane, ``Optimal {DR} and {ESS} scheduling for
  distribution losses payments minimization under electricity price
  uncertainty,'' \emph{IEEE Transactions on Smart Grid}, vol.~7, no.~1, pp.
  261--272, 2016.

\bibitem{2023:Vignesh}
V.~Ramasamy, J.~Zuboy, D.~Feldman, R.~Margolis, J.~Desai, A.~Walker,
  M.~Woodhouse, E.~O'Shaughnessy, and P.~Basore, ``{Q1} 2023 {U.S.} solar
  photovoltaic system and energy storage cost benchmarks with minimum
  sustainable price analysis data file,'' \emph{NREL Data Catalog. Golden, CO:
  National Renewable Energy Laboratory}, 2023.

\bibitem{2017:Raoof}
R.~Hasanpour, B.~M. Kalesar, J.~B. Noshahr, and P.~Farhadi, ``Reconfiguration
  of smart distribution network considering variation of load and local
  renewable generation,'' in \emph{2017 IEEE International Conference on
  Environment and Electrical Engineering and 2017 IEEE Industrial and
  Commercial Power Systems Europe}, 2017, pp. 1--5.

\end{thebibliography}

\end{document}